\documentclass[         %
superscriptaddress,    
nofootinbib,            
onecolumn,             %
floatfix]               
{revtex4-1}               

\usepackage{amssymb}
\usepackage{graphicx}
\usepackage{amsmath}
\usepackage[hypertex]{hyperref}
\usepackage{amsfonts}
\usepackage{natbib}
\usepackage{lipsum}

\usepackage{tabularx}
\usepackage{xcolor}
\usepackage{oubraces}
\usepackage{dsfont}
\usepackage{titlesec}
\usepackage{transparent}
\usepackage{tocloft}
\usepackage{color}

\begin{document}
\title{Emergence of correlated proton tunneling
in water ice}

\author{Onur Pusuluk}
\affiliation{Department of Physics, Ko\c{c} University, Sar{\i}yer,
\.{I}stanbul, 34450 Turkey} \affiliation{Department of Physics,
\.{I}stanbul Technical University, Maslak, \.{I}stanbul, 34469
Turkey}
\author{Tristan Farrow}
\affiliation{Department of Physics, University of Oxford, Parks
Road, Oxford, OX1 3PU, UK} \affiliation{Centre for Quantum
Technologies, National University of Singapore, 3 Science Drive 2,
Singapore 117543, Singapore}
\author{Cemsinan Deliduman}
\affiliation{Department of Physics, Mimar Sinan Fine Arts
University, Bomonti, \.{I}stanbul, 34380, Turkey}
\author{Vlatko Vedral}
\affiliation{Department of Physics, University of Oxford, Parks
Road, Oxford, OX1 3PU, UK} \affiliation{Centre for Quantum
Technologies, National University of Singapore, 3 Science Drive 2,
Singapore 117543, Singapore}

\date{\today}

\begin{abstract}

Several experimental and theoretical studies report instances of
concerted or correlated multiple proton tunneling in solid phases of
water. Here, we construct a pseudo-spin model for the quantum motion
of protons in a hexameric H$_2$O ring and extend it to open system
dynamics that takes environmental effects into account in the form
of O$-$H stretch vibrations. We approach the problem of correlations
in tunneling using quantum information theory in a departure from
previous studies. Our formalism enables us to quantify the coherent
proton mobility around the hexagonal ring by one of the principal
measures of coherence, the $l_1$ norm of coherence. The nature of
the pairwise pseudo-spin correlations underlying the overall
mobility is further investigated within this formalism. We show that
the classical correlations of the individual quantum tunneling
events in long-time limit is sufficient to capture the behaviour of
coherent proton mobility observed in low-temperature experiments. We
conclude that long-range intra-ring interactions do not appear to be
a necessary condition for correlated proton tunneling in water ice.


\end{abstract}

\maketitle

\section{Introduction} \label{Intro}

Hydrogen bonding (or H-bonding) is the subject of extensive
literature due to its central importance in many natural phenomena
in physical, chemical, and biological systems. The first attempts
that reveal the underlying physics behind it go back to the 1950s
\cite{1959_Sokolov, 1960_Pauling}, and since then, quantum aspects
of the nature of this weak interaction are still being hotly
debated. In the meantime, most of the demystification attempts have
focused on the H-bonds using water as an explanatory model
\cite{2011_RevCovOFHB, 2015_CovOfHBInLiquidWater, 2012_McKenzie,
2014_McKenzie}.

Let's designate a H-bonded system X$_1-$H$\cdot\cdot\cdot$X$_2$
where the single covalent bond X$_1-$H is a proton-donating bond,
X$_1$ is the proton-donor and X$_2$ is the proton-acceptor. One
controversial issue about the role of non-trivial quantum effects in
such a system is the extent of the covalency of the
H$\cdot\cdot\cdot$X$_2$ interaction \cite{2011_RevCovOFHB}, i.e.,
charge transfer from the lone pair orbital of the proton-acceptor
($\sigma_{LP(\text{X$_2$})}$) to the unoccupied antibonding orbital
of the proton-donating bond ($\sigma^{\ast}_{\text{X$_1-$H}}$).
Although the covalent contribution to the attractive energy of
H-bonds in water is comparable to the electrostatic contribution,
the amount of charge transfer itself is of the order of just a few
millielectrons \cite{2015_CovOfHBInLiquidWater}.

Apart from this intermolecular charge transfer, non-trivial quantum
effects also enter into the physics of H-bonding in the form of
proton tunneling back and forth between donor and acceptor.
According to diabatic state models \cite{2012_McKenzie,
2014_McKenzie}, nuclei of H atoms are likely to tunnel through
H-bonds between water monomers. Several \textit{ab inito} studies
examined this probability in water ice as well. First and foremost,
proton tunneling was found to be responsible for the pressure driven
phase transitions from proton-ordered ice VIII to proton-disordered
ice VII around $100$ K \cite{1998_PTInIceX}, and is believed to
drive the transition from proton-disordered ice I$_\text{h}$ to
proton-ordered ice XI in a microscopic model \cite{2006_PTInIceIh}.

However, spontaneous single proton tunnelings violate the so-called
Bernal-Fowler \textit{ice rules} \cite{1933_BF, 1935_Pauling} which
state that (i) each water molecule is linked to four other ones
through H-bonds in such a way that (ii) it behaves as a proton-donor
in half of these four bonds and a proton-acceptor in the remaining
ones. These local constraints are expected to lead to correlations
between individual proton tunnelings. Consistent with this
expectation, the likelihood of correlated proton tunneling in water
ice was recently reported by successive low-temperature experiments
such as incoherent quasielastic neutron scattering measurements on
ice I$_\text{h}$ and I$_\text{c}$ \cite{2009_CorrPTInIceIhAndIc},
scanning tunnelling microscopy of cyclic water tetramer
\cite{2015_CorrPTInTatramer}, and high precision measurements of the
complex dielectric constant of ice XI \cite{2015_CorrPTInIceXI}.
Additionally, the trace of correlations of the individual quantum
tunneling events in water ice has been theoretically explored using
several models. One-particle density matrix analysis confirmed the
presence of proton correlations in ice VII, but not in ice VIII and
I$_\text{h}$ \cite{2011_CorrPTInIceVII}. On the contrary, concerted
tunneling of six protons in ice I$_\text{h}$ appeared to occur at
low temperatures in path integral simulations
\cite{2014_CorrPTInIceIh} and in lattice-based calculations
\cite{2016_CorrPTInIceIh}.

Here, we introduce a pseudo-spin model for the quantum motion of
protons in a hexameric H$_2$O ring. Unlike considerations of
pseudo-spins by previous studies \cite{2006_PTInIceIh,
2016_CorrPTInIceIh}, we do not attempt to impose the collective six
proton tunneling by effectively incorporating a single matrix
element into the Hamiltonian, or to map the problem onto a lattice
gauge theory. Instead, we develop an extension of the model to open
system dynamics and approach the correlation problem from the
standpoint of quantum information theory. Temperature dependence of
proton correlations in (athermal) equilibrium are monitored by
well-known measures of quantumness such as $l_1$ norm of coherence
\cite{Plenio-2014}, relative entropy of coherence
\cite{Plenio-2014}, entanglement of formation \cite{Wooters-1996},
concurrence \cite{Wooters-2000}, quantum discord \cite{Vedral-2001,
Zurek-2002}, and geometric measure of discord \cite{Vedral-2010}.
\begin{figure}[h] \centering
        \begin{minipage}[t]{0.3\linewidth}
        \includegraphics[width=\linewidth]{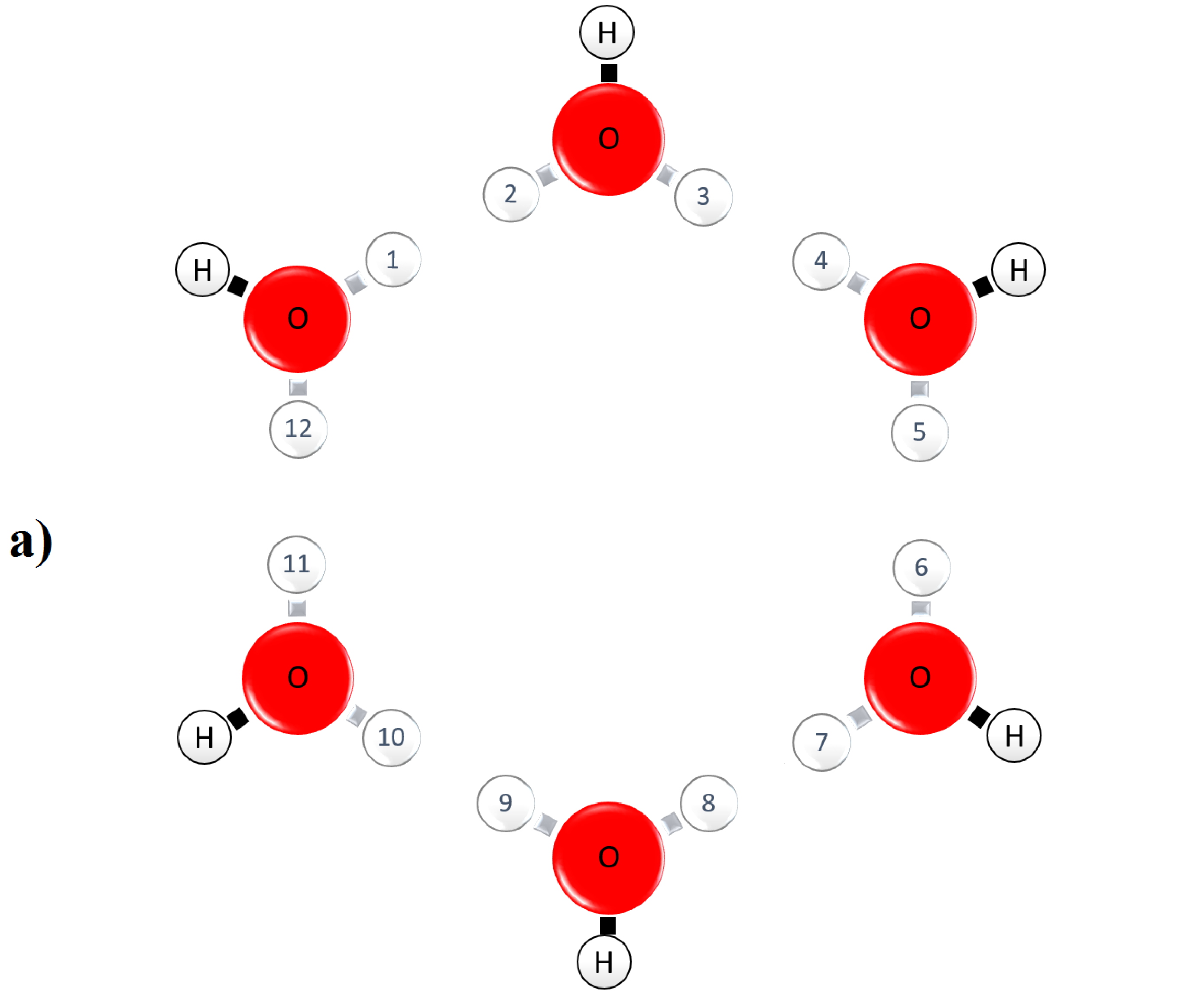}
        \end{minipage}
        \begin{minipage}[t]{0.3\linewidth}
        \includegraphics[width=\linewidth]{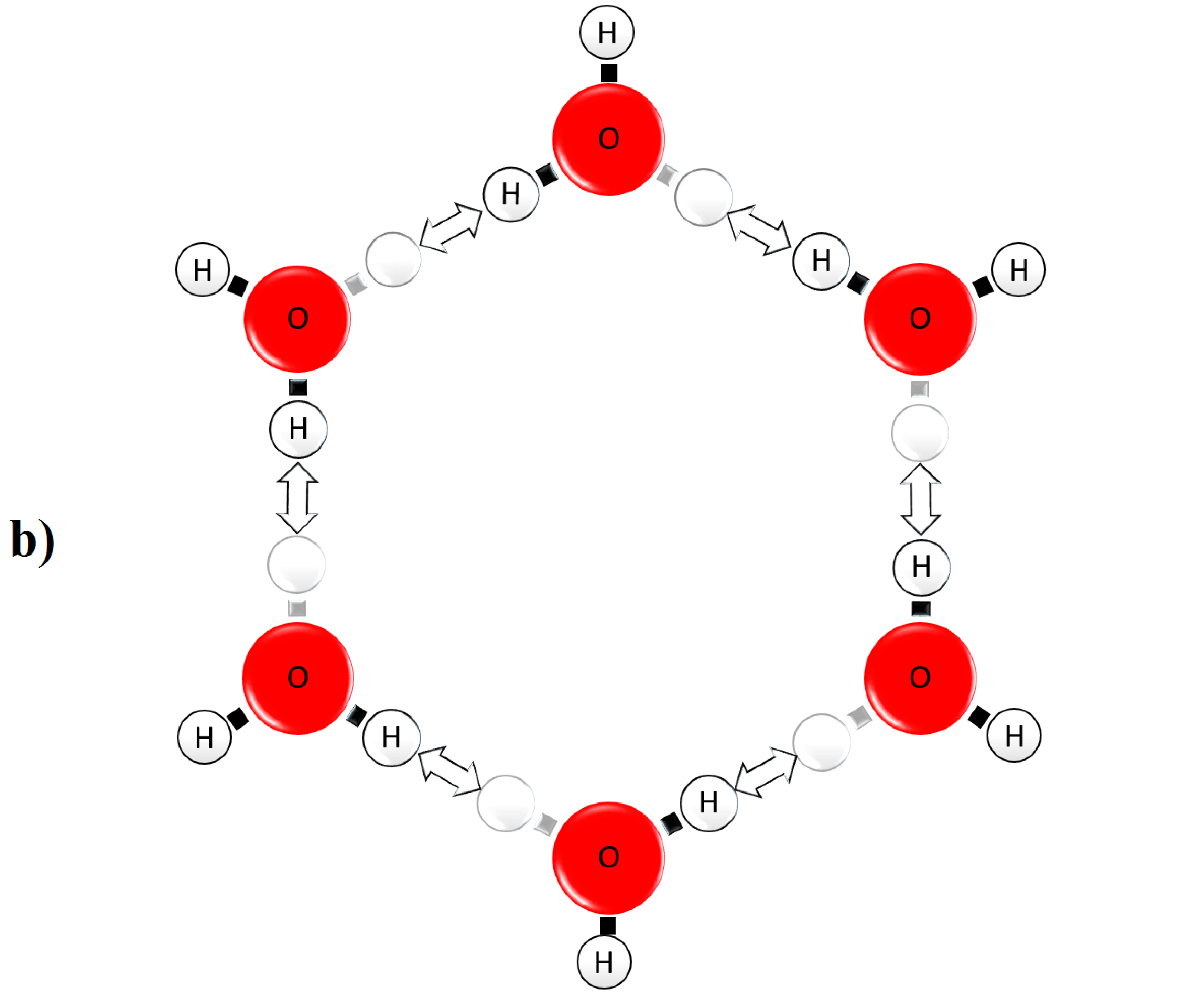}
        \end{minipage}
        \begin{minipage}[t]{0.3\linewidth}
        \includegraphics[width=\linewidth]{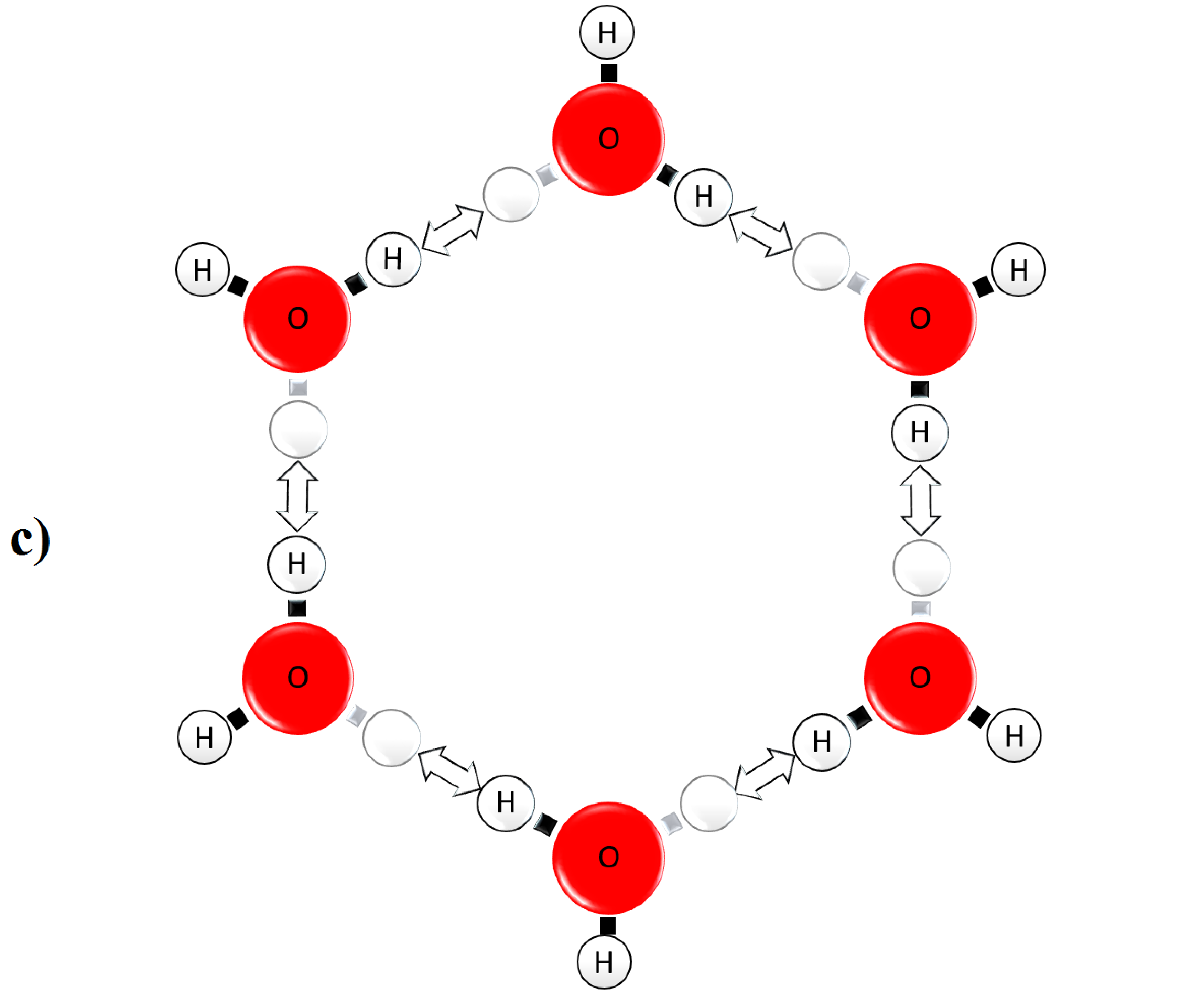}
        \end{minipage}
        \caption{(a) The hexagonal lattice. It is made of OH$^-$ ions
        in fixed positions relative to each other. Edges can be interpreted
        as H-bonds. Protons are allowed to live on the enumerated locations.
        Ice rules are satisfied only by $2$ of the six-proton configurations given in (b) and (c).
        Transitions between them require simultaneous
        relocations of the six protons within the ring in the directions depicted
        by two-sided arrows. Any other configuration accessible from (b) or (c)
        by successive proton relocations in H-bonds is called an ionic defect.
        Bjerrum defects occur in all the other configurations including
        the remaining six-proton configurations.}
        \label{Model}
\end{figure}

\section{Model and Methods} \label{Model_Part1}

Although their unit cells belong to different space groups, the
basic structures of both ice I$_\text{h}$ and XI can be visualized
as a hexameric box whose planes are either chair-form or boat-form
3-d hexamers. To reduce the complexity, we restrict our model to a
2-d hexagonal ring with a hydroxyl ion (OH$^-$) resides in each
vertex, as shown in figure \ref{Model}-a. Rigid rotations of the
vertices are not taken into account because of the high energy cost
assumed in microscopic models \cite{2006_PTInIceIh} and predicted in
experiments \cite{2015_CorrPTInIceXI}. Each edge linking two
vertices represents a H-bond and includes two equally likely
locations for H$^+$ ions. These locations (enumerated in figure
\ref{Model}-a) can be regarded as a crystal lattice in which H$^+$
ions, or protons, move according to the Hamiltonian
\begin{eqnarray}\label{Ham_HL0} \begin{aligned}
H_{H\!ex} = \sum_{j=1}^{12} W_j n_j - \sum_{j=1}^{12} J_{j,j+1}
(a_j^{\dagger} a_{j+1} + a_j a_{j+1}^{\dagger}) + \sum_{j=1}^{12}
V_{j,j+1} n_j n_{j+1} + \lambda ,
\end{aligned}
\end{eqnarray}
where subscripts are in mod $12$, $n_{j} = a_{j}^{\dagger} a_{j}$ is
the proton number operator at lattice site $j$, $a_{j}^{\dagger}$
and $a_{j}$ are respectively proton creation and annihilation
operators that obey the following anticommutation relations
\begin{eqnarray} \label{anticomm}
\{ a_{j}, a_{k} \} = \{ a_{j}^{\dagger}, a_{k}^{\dagger} \} = 0, \{
a_{j}, a_{k}^{\dagger} \} = \delta_{j k} .
\end{eqnarray}

On-site energy $W_{j}$ can be taken as the total potential felt by a
proton at $j$th site, i.e., the sum of a Morse potential describing
the single (covalent) bond with the adjacent OH$^-$ ion and a
Coulomb potential representing the electrostatic attraction to the
opposite OH$^-$ ion. $J_{j,j+1}$ stands for the orbital interactions
which causes proton tunneling. Its expected dependence on the
geometry implies that $J_{j,j+1}^{(\text{edge})} \gg
J_{j,j+1}^{(\text{vertex})}$ where the former is the intermolecular
proton tunneling coefficient for the neighbouring sites occupying
the same edge, while the latter is the intramolecular proton
tunneling coefficient for the successive sites close to the same
vertex. As $J_{j,j+1}^{(\text{edge})}$ should already be quite small
compared to other coefficients, we can neglect
$J_{j,j+1}^{(\text{vertex})}$. This assumption guarantees the
absence of any quantum correlation between the quantum tunneling
events through individual hydrogen bonds in the isolated hexamer.
$V_{j,j+1}$ is introduced to penalize two-proton cases associated
with the violation of ice rules. Edge sharing sites and vertex
sharing sites have different penalty coefficients as well as
different tunneling coefficients. Presence of two protons on the
same edge is called as a Bjerrum defect, whereas occupation of both
sites near the same vertex is called as an ionic defect. As Bjerrum
defects require more energy, $V_{j,j+1}^{(\text{edge})} \gg
V_{j,j+1}^{(\text{vertex})}$. Finally, $\lambda$ is a constant
responsible for the total intermolecular interactions between
vertices, such as Pauli repulsion, Van der Walls interaction, and
London dispersion.

Symmetry of the lattice provides that $W_{j} = W$,
$J_{j,j+1}^{(\text{edge})} = J$, $V_{j,j+1}^{(\text{edge})} =
V_{inter}$, and $V_{j,j+1}^{(\text{vertex})} = V_{intra}$.

So far, we have focused only on nearest-neighbor interactions and
neglected the further interactions with other neighbors, including
the concerted tunneling of the six protons arising from the
collective overlap of the orbitals. The effect of long-range
interactions on the proton dynamics in water ice was in fact
proposed to be negligible at low temperatures \cite{2006_PTInIceIh}.
However, unlike long-range intra-ring interactions, the long-range
inter-ring interactions are expected to be non-negligible for the
single-ring dynamics. The electrostatic and topological interactions
with adjacent rings should at least have significant effects on the
parameters $W$, $J$, $ V_{inter}$, and $V_{intra}$. We assume that
the effects of other rings on each of these parameters can be
approximated respectively by a single averaged effect. In what
follows $W$, $J$, $ V_{inter}$, and $V_{intra}$ are recounted as
effective parameters that include mean-field averages.

To obtain a pseudo-spin Hamiltonian by preserving the
anti-commutation relations (given in equation (\ref{anticomm})), we
apply the Jordan-Wigner transformation for $a_{j}$,
$a_{j}^{\dagger}$, and $n_{j}$ in equation (\ref{Ham_HL0}) in such a
way below
\begin{eqnarray}\begin{aligned} \label{Jordan-Wigner}
a_{j} &= \exp \left( - \text{i} \pi \sum_{k=1}^{j-1} \sigma_-^{(k)}
\sigma_+^{(k)} \right) \sigma_+^{(j)} , \\
a_{j}^{\dagger} &= \exp \left( + \text{i} \pi \sum_{k=1}^{j-1}
\sigma_-^{(k)}
\sigma_+^{(k)} \right) \sigma_-^{(j)} , \\
n_{j} &= \sigma_-^{(j)} \sigma_+^{(j)} ,
\end{aligned}
\end{eqnarray}
where $\sigma_-^{(j)} = |1_{j}\rangle \langle0_{j}|$ and
$\sigma_+^{(j)} = |0_{j}\rangle \langle1_{j}|$ with the convention
for Pauli $z$ operator that $\sigma_z^{(j)} = |0_{j}\rangle
\langle0_{j}| - |1_{j}\rangle \langle1_{j}|$. Note that in contrast
to the standard application of the Jordan-Wigner transformation on
electron transport phenomena, the creation of a proton at the $j$th
site is an energy lowering process here.

After writing (\ref{Jordan-Wigner}) in terms of Pauli operators,
i.e., $n_{j} = \frac{1}{2} \left( I^{(j)} - \sigma_z^{(j)} \right)$
and $\sigma_\pm^{(j)} = \frac{1}{2} \left(\sigma_x^{(j)} \pm
\text{i} \sigma_y^{(j)} \right) $, we substitute it into
(\ref{Ham_HL0}) and arrive at the following pseudo-spin Hamiltonian
\begin{eqnarray} \label{Ham_HL} \begin{aligned}
H_{H\!ex} = &+ \sum_{j=1}^6 J_x \left(\sigma^{(2j-1)}_{x} \otimes
\sigma^{(2j)}_{x} + \sigma^{(2j-1)}_{y} \otimes
\sigma^{(2j)}_{y}\right) \\
&+ \sum_{j=1}^6 J_z^{(\text{inter})} \left( \sigma^{(2j-1)}_{z}
\otimes
\sigma^{(2j)}_{z} \right) \\
&+ \sum_{j=1}^6 J_z^{(\text{intra})} \left(
\sigma^{(2j)}_{z} \otimes \sigma^{(2j+1)}_{z} \right) \\
&+ \sum_{j=1}^{12} B \, \sigma^{(j)}_{z} + \tilde{\lambda} ,
\end{aligned}
\end{eqnarray}
where the superscripts of the Pauli matrices are in mod $12$, $J_x =
- J/2$, $J_z^{(\text{inter})} = V_{inter}/4$, $J_z^{(\text{intra})}
= V_{intra}/4$, $B = - (2\,W + V_{inter} + V_{intra})/4$, and
$\tilde{\lambda} = \lambda + 6 \, W + 3 \, (V_{inter} +
V_{intra})/2$. Note that these parameters have some contributions
from the mean-field averages of the effects of the surrounding
hexamers and the construction of the Hamiltonian guarantees the
absence of any quantum correlation between the quantum tunneling
events through individual hydrogen bonds in the isolated hexamer.
Also note that the same pseudo-spin formalism described above has
been recently used in \cite{2018_Pusuluk} to investigate the role of
proton tunneling in biological catalysis.

The most general quantum state of pseudo-spins can be described
using density matrix formalism such that each computational basis
state represents a different configuration of protons. For example,
configurations in figures \ref{Model}-b and c are respectively
represented by basis states $| 010101010101 \rangle$ and $|
101010101010 \rangle$, so that the protons can exist in any coherent
(or  incoherent) superposition of these states during the dynamical
evolution of the closed (or open) system. Suppose that the proton
residing at the site $j = 2$ in the former configuration moves to
the site $j = 1$ by either classical hopping or quantum tunneling
between the initial and final times $t_i$ and $t_f$. In the case of
thermally activated classical hopping, it leaves the site $j = 2$ at
$t = t_i$, then enters into the exterior space between the sites,
and finally reaches the site $j  = 1$ at $t = t_f$. Since we do not
take into account its presence in the exterior space between the
sites, it disappears at the site $j = 2$ at $t = t_i$ and reappears
at the site $j  = 1$ at $t = t_f$ in our model, i.e., $ \rho(t) = |
0 \rangle \langle 0 |$ where $| 0 \rangle \equiv | 000101010101
\rangle$ and $t_i < t < t_f$. Conversely, it will never enter into
the exterior space between the sites in the course of its motion
during tunneling, but it will be delocalized between both sites,
i.e., $\rho(t) = | \psi \rangle \langle \psi |$ where $|\psi \rangle
= \alpha | 1 \rangle + \beta | 2 \rangle$, $| 1 \rangle \equiv |
100101010101 \rangle$, $| 2 \rangle \equiv | 010101010101 \rangle$,
$|\alpha|^2 + |\beta|^2 = 1$, and $t_i < t < t_f$. In this case, it
is found at the first site with a probability of $|\alpha|^2$ if its
location is measured at any time between $t_i$ and $t_f$. The
decoherence taking place at the end of quantum tunneling, at $t =
t_f$, converts the coherent superposition $| \psi \rangle \langle
\psi |$ into incoherent superposition $|\alpha|^2 | 1 \rangle
\langle 1 | + |\beta|^2 | 2 \rangle \langle 2 |$ removing the
off-diagonal elements from the density matrix.

The unidirectional tunneling process described above can be captured
by a snapshot of the state within our pseudo-spin formalism, that is
to say, we can detect it by evaluating the off-diagonal elements of
the pseudo-spin density matrix at a single time. The advantages of
our approach go beyond this. Consider the static structures of
pseudo-spins, e.g., the ground state during closed-system dynamics
or the thermal state during open-system dynamics. When such a
stationary state equals to $| \psi \rangle \langle \psi |$, it means
that the proton under consideration delocalizes between the first
and second sites over a long time period. This can be regarded as a
quantum tunneling of the proton back and forth between these two
sites since the state never collapses onto the basis states, which
represent the proton localizations at their respective sites (see
the same usage of the term in \cite{2015_CorrPTInIceXI,
2011_CorrPTInIceVII}) . In contrast, we cannot claim a back and
forth tunneling event when the steady state is found to be
$|\alpha|^2 | 1 \rangle \langle 1 | + |\beta|^2 | 2 \rangle \langle
2 |$. Hence, in addition to unidirectional proton tunneling events
in the short time limit, the bidirectional proton tunneling events
in the long time limit are also described by well-defined density
matrices.

Mixed states like the incoherent superpositions above cannot be
generated from pure initial states during the closed-system dynamics
governed by the Hamiltonian (\ref{Ham_HL}). But they can be
generated from coherent superposition states as a result of
environmental decoherence that will be described in what follows.

\subsection{Open system dynamics}

It is hard to draw a generic model of the environment for the motion
of protons through H-bonds. Such a model should include at least
three kinds of vibrations as each individual H-bond is defined by
three geometric parameters, e.g., length of the proton-donating
bond, donor-acceptor separation and bond angle. However, a
minimalistic model consisting of just the periodic oscillations
associated with the lengths of proton-donating bonds seems to be
sufficient to describe the low-temperature dynamics of protons in a
hexameric H$_2$O ring as in the following. These oscillations can be
incorporated into our model as independent thermal baths around
lattice sites with individual self-Hamiltonians
\begin{eqnarray}\label{HamB}
H_B^{(j)} = \sum_k \hbar \omega_{j, k} \, b_{j, k}^{\dagger} b_{j,
k} ,
\end{eqnarray}
where $b_{j, k}^{\dagger}$ and $b_{j, k}$ are phonon creation and
annihilation operators associated with the $k$th oscillator mode at
the $j$th site. We assume that the equilibrium positions of the
protons are linearly coupled to the positions of the phonons through
\begin{eqnarray}\begin{aligned} \label{HamI1}
H_{I} &= \sum_{j} n_{j} \,\,\, \sum_k  \, \left(g_{j, k} b_{j,
k}^{\dagger} + g^*_{j, k} b_{j, k}\right)
\\
&\propto \sum_{j} \sigma_z^{(j)} \sum_k  \, \left(g_{j, k} b_{j,
k}^{\dagger} + g^*_{j, k} b_{j, k}\right) .
\end{aligned}
\end{eqnarray}

It is important to realize that the local interaction described
above induces the entanglement of each pseudo-spin with the
positions of associated phonons. In the absence of spin-spin
coupling ($J_x = 0$), the dynamics of the pseudo-spins are fully
separated from each other, and each pseudo-spin undergoes a pure
dephasing process.

Before extending this discussion to the case of non-vanishing
inter-spin coupling, let us first examine the role of memory effects
in open system dynamics. The Born-Markov approximation can be
justified only if the state of pseudo-spins varies over a time scale
much longer than the lifetime of the environmental excitations. The
vibration of the O$-$H bond in O$-$H$\cdot\cdot\cdot$O systems has a
period of $\simeq 10$ fs, which corresponds to a stretch harmonic
frequency of $\simeq 3500$ cm$^{-1}$. Unlike the short-lived
($\approx 1$ ps) H-bonds in liquid water \cite{2001_HBLifeTime},
H-bonds survive sufficiently long in ice I$_\text{h}$ and the jump
time of protons in these bonds is larger than tens of fs, e.g., is
equal to $3.7$ ps at $5$ K \cite{2009_CorrPTInIceIhAndIc}. So, we
can describe the picosecond evolution of the pseudo-spins' state
$\rho$ on the basis of a Markovian master equation in the following
Lindblad form \cite{BreuerAndPetruccione-2002}
\begin{eqnarray}\label{MasterEq}
\frac{d\rho}{dt} = - \frac{\text{i}}{\hbar} [H_{H\!ex} + \hbar
H_{L\!S},\rho] + \mathcal{D}(\rho) ,
\end{eqnarray}
where the Lamb shift Hamiltonian provides a unitary contribution to
the open dynamics and reads
\begin{eqnarray}\label{LambShift}
H_{L\!S} = \sum_{\omega} \sum_{j, j^{\prime}} S_{j
j^{\prime}}(\omega) \, A_{j}^{\dagger}(\omega)\,
A_{j^{\prime}}(\omega) ,
\end{eqnarray}
whereas the dissipator is defined by
\begin{eqnarray} \begin{aligned} \label{Dissip}
\mathcal{D}(\rho) = \sum_{\omega} \sum_{j,j^{\prime}} \gamma_{j
j^{\prime}}(\omega) \, \big(A_{j^{\prime}}(\omega) \rho
A_{j}^{\dagger}(\omega) - \frac{1}{2} \{A_{j}^{\dagger}(\omega)\,
A_{j^{\prime}}(\omega), \rho \} \big) , \end{aligned}
\end{eqnarray}
with $\omega = \epsilon_m - \epsilon_{m^{\prime}}$. Here,
$\epsilon_m$'s are the eigenvalues of pseudo-spin Hamiltonian given
in (\ref{Ham_HL}) and Noise operators $A_{j}(\omega)$ are the
eigenoperators of this self-Hamiltonian
\begin{eqnarray}\label{NoiseOp}
A_{j}(\omega) = \sum_{\epsilon_m - \epsilon_{m^{\prime}} = \omega}
|\epsilon_{m^{\prime}}\rangle \langle\epsilon_{m^{\prime}}|
A_{\alpha} |\epsilon_m\rangle \langle\epsilon_m| ,
\end{eqnarray}
where $A_{j}$ are the Hermitian operators coupled to the
environment, i.e., Pauli $z$ operators as introduced in equation
(\ref{HamI1}). Coefficients $S_{j j^{\prime}}(\omega)$ and
$\gamma_{j j^{\prime}}(\omega)$ are respectively the imaginary part
and half of the real part of the one-sided Fourier transformation of
the thermal bath correlation function given by
\begin{eqnarray}\label{MasterEqGamma}
\Gamma_{j j^{\prime}}(\omega) = \frac{1}{\hbar^2} \int_0^{\infty} ds
\, e^{i \, \omega \, s} \left\langle B_{j}^{\dagger}(s)
B_{j^{\prime}}(0) \right\rangle_{\text{th}} ,
\end{eqnarray}
where $ B_{j}(s)$ are the interaction picture representations of the
bath operators (included in (\ref{HamI1})) and $J(\omega)$ is the
spectral density function encapsulating all the effects of the
environment. We assume that each pseudo-spin is associated with an
independent environment, $\Gamma_{j j^{\prime}} = \Gamma_{j j} \,
\delta_{j j^{\prime}}$. Furthermore, we focus on a symmetric lattice
at a constant temperature $T$ that makes these individual baths
identical, so  $\Gamma_{j j} = \Gamma$.

The normal modes of lattice vibrations are more complicated in real
water ice. The lattice sites, especially the pair of sites sharing
the same vertex, are so close to each other that it is expected to
find correlations between them. Presence of the correlations between
the individual baths may result in the emergence of quantum
correlations between the quantum tunneling events in the course of
open system dynamics. However, we would like to restrict our
analysis to the importance of classical correlations between the
quantum tunneling events on the overall proton mobility. Hence, we
ignore the correlations between the individual baths as well as the
Hamiltonian parameter $J_{j,j+1}^{(\text{vertex})}$.

\subsection{Measures of quantum correlations} \label{QIT}

Quantum coherence is the degree of quantum superposition found in a
generic state $\rho$ with respect to a given orthogonal basis
$\{|m\rangle\}$. One of the most widely used measures that satisfy
all the requirements for a proper measure of quantum coherence is
the $l_1$ norm of coherence \cite{Plenio-2014}, defined as
\begin{equation} \label{Eq_l1Norm}
C_{l_1}\!\left[\rho\right] = \sum_{m \neq m^\prime} | \langle m
|\rho|m^\prime\rangle | .
\end{equation}

Off-diagonal elements of the density matrix $\rho$ are related to
the transitions between computational basis states, and each basis
state represents a different configuration of the protons in our
model. Hence, the $l_1$ norm of the pseudo-spins' state quantifies
the quantum coherent proton mobility when the proton number is
fixed. As an example, $|m\rangle$ equals to $|b\rangle \equiv | 01
\rangle^{\bigotimes 6}$ and $|c\rangle \equiv | 10
\rangle^{\bigotimes 6}$ for the configurations respectively depicted
in figures \ref{Model}-b and c. A transition from one of these
states to the other requires simultaneous relocations of the six
protons to their adjacent empty sites in H-bonds. Any quantum
coherent superposition $|\psi\rangle = \alpha |b\rangle + \beta
|c\rangle$ with $|\alpha|^2 + |\beta|^2 = 1$ represents such a
motion if it takes place in the form of concerted tunneling of six
protons between the corresponding configurations. The extent of the
quantum character of this motion of the protons is reflected in the
density matrix by the off-diagonal elements $\langle b  | \rho | c
\rangle = \alpha \beta^*$ and $\langle c  | \rho | b \rangle =
\alpha^* \beta$. The sum of the absolute values of these elements is
maximum when $\alpha = \beta = 1/\sqrt{2}$, and this corresponds to
a quantum state in which the protons can be found in each of the two
configurations with a probability of $1/2$ if their locations are
measured.

Another proper measure of coherence is the relative entropy of
coherence \cite{Plenio-2014}:
\begin{equation}
C_{R}^{\mathrm{IC}}\!\left[\rho\right] = \min_{\varsigma \in
\mathrm{IC}} \left(S\left[\rho || \varsigma\right]\right) =
S[\rho_d] - S[\rho] ,
\end{equation}
where the minimum is taken over the set of incoherent states (IC)
that are diagonal in the basis $\{|m\rangle\}$, $S\left[\rho ||
\varsigma\right]$ is the quantum relative entropy that equals to $-
\mathrm{tr}\!\left[\rho \, \left(\log_2 \rho - \log_2
\varsigma\right)\right]$, $S[\rho]$ is the von Neumann entropy
 that equals to $- \mathrm{tr}\!\left[\rho \, \log_2 \rho\right]$, and
$\rho_d$ is the diagonal part of the density matrix $\rho$. That is
to say, $C_{R}^{\mathrm{IC}}\!\left[\rho\right]$ measures the
distinguishability of a density matrix with a modified copy in which
the off-diagonal elements are removed by a full dephasing process.
Whereas $C_{l_1}\!\left[\rho\right]$ takes into account distinct
tunneling pathways independently of each other,
$C_{R}^{\mathrm{IC}}\!\left[\rho\right]$ does not discriminate
between these pathways that rearrange the proton configuration and
highlights the overall nonclassical mobility, hence showing a
holistic picture.

Quantum correlations also arises from the superposition principle.
Nonlocal correlations found in nonseparable quantum superposition
states are known as quantum entanglement. Entanglement of formation
\cite{Wooters-1996} is a good measure of entanglement for a generic
bipartite state $\rho$ and is defined as
\begin{equation} \label{Eq_EoF1}
E_F\!\left[\rho\right] = \min \left( \sum_i \mathrm{Q}^{(i)}
E_E\!\left[| \psi_i \rangle \langle \psi_i |\right] \right) ,
\end{equation}
where the minimum is taken over all the possible pure state
decompositions that realize $\rho = \sum_i \mathrm{Q}^{(i)} | \psi_i
\rangle \langle \psi_i | $, and $E_E$ is the entropy of
entanglement, the unique measure of entanglement for pure bipartite
states that equals to the von Neumann entropy of the reduced state
of one of the two subsystems, i.e., $E_E[\rho] = S[\rho_{1(2)}] = (S
\circ \mathrm{tr}_{2(1)})[\rho]$. Although it is hard to compute
$E_F\!\left[\rho\right]$ for a general state, even numerically, an
explicit formula in the form of a binary entropy can be derived in
the case of two-qubit systems:
\begin{equation} \label{Eq_EoF2}
E_F\!\left[\rho\right] = - f \log_2 ( f ) - ( 1 - f ) \log_2 ( 1 - f
) ,
\end{equation}
where $f = f(C) = (1 + \sqrt{1 - C^2})/2$ and $C$ is an entanglement
monotone, called concurrence \cite{Wooters-2000}. This entanglement
monotone is defined as
\begin{equation} \label{Eq_Conc}
C[\rho] = \max \left(0,\sqrt{\lambda_1} - \sqrt{\lambda_2} -
\sqrt{\lambda_3} - \sqrt{\lambda_4}\right) ,
\end{equation}
where $\lambda_i$ are the eigenvalues of the operator $\rho
(\sigma_{y} \otimes \sigma_{y}) \, \rho^{\ast} (\sigma_{y} \otimes
\sigma_{y})$ in decreasing order. In addition to providing an
explicit formula for the entanglement of formation, the concurrence
can also be used to reveal one of the most fundamental properties of
entanglement known as the monogamy of entanglement
\cite{Wooters-2000}, which imposes a trade-off between the amount of
entanglement between different subsystems in a composite system.

As each edge in figure \ref{Model}-a represents a H-bond, the
entanglement of formation of the reduced state of pseudo-spins lying
on the same edge quantifies the entanglement generated by proton
tunneling through the corresponding H-bond. Besides this, non-zero
entanglement of formation between a pseudo-spin pair lying on
different edges indicates the inter-bond entanglement between the
protons that belong to corresponding H-bonds. The trade-off between
intra-bond and inter-bond entanglements can then be investigated
using the concurrence. However, quantum correlations are not limited
to quantum entanglement, e.g., separable mixed states can possess
nonclassical correlations known as quantum discord
\cite{Vedral-2001, Zurek-2002} when the orthogonality condition on
local bases breaks down at least in one subsystem and provides local
indistinguishability. The original definition of quantum discord
\cite{Zurek-2002} is grounded in the difference between two
different quantum generalizations of mutual information:
\begin{eqnarray} \begin{aligned} \label{Eq_qDiscord1}
\delta_{1:2} &= I[\rho] - J[\rho] \equiv \left(- S[\rho]
+ S[\rho_{1}] + S[\rho_{2}]\right) - \max_{\{M_{i}^{(2)}\}}
\left(S[\rho_1] - S[\rho_1|\{M_{i}^{(2)}\}]\right) \\
&= \min_{\{M_{i}^{(2)}\}} \left(S[\rho_2] + \sum_{i} \mathrm{P}_{i}
S[\rho_{1|M_{i}^{(2)}}] - S[\rho] \right) ,
\end{aligned} \end{eqnarray}
where $S[\rho_1|\{M_{i}^{(2)}\}]$ is the quantum conditional entropy
of the first subsystem, given the complete measurement
$\{M_{i}^{(2)}\}$ on the second subsystem and $\rho_{1|M_{i}^{(2)}}
= \mathrm{tr}_{2} [M_i \, \rho \, M_i^\dagger]/\mathrm{P}_{i}$ are
post-measurement states of the first subsystem with corresponding
probabilities $\mathrm{P}_{i} = \mathrm{tr}_{2} [M_i^\dagger \, M_i
\, \rho]$.

Quantum discord is the measure of nonclassical correlations which
includes entanglement as a subset since the first quantum
generalization of mutual information measures the total correlations
between subsystems, whereas the second generalization $J[\rho]$
quantifies only classical correlations \cite{Vedral-2001}. In other
words, quantum information theory can describe all correlations
contained in a complex system in a rigorous way. Classical
correlations generated by classical proton hopping through a H-bond
can be quantified by the mutual information $J$ between the
pseudo-spins lying on the edge corresponding to this bond. Moreover,
classical correlations between a pair of protons that belong to
different H-bonds in the H$_2$O hexamer can be captured by
$J[\rho_{j j^\prime}]$ where $\rho_{j j^\prime}$ is the reduced
state of pseudo-spins $j$ and $j^\prime$ that lie on the edges
corresponding to these H-bonds. Note that the classical motion of
protons does not appear during the closed system dynamics in our
model, but arises from the interaction with the environment.

Since any analytic expression is unknown for the mutual information
based measure $\delta_{1:2}$ in a generic system, it is hard to
evaluate it. However, an explicit formula of a distance-based
measure known as the geometric measure of discord \cite{Vedral-2010}
is available for a generic two-qubit system which can be written in
the Bloch representation:
\begin{eqnarray} \begin{aligned}
\rho &= \frac{1}{4} \Big( \mathbb{I}_1 \otimes \mathbb{I}_2 +
\sum_{i=1}^3 x_i \, \mathbb{I}_1 \otimes \sigma_i^{(2)} +
\sum_{i=1}^3 y_i \, \sigma_i^{(1)} \otimes \mathbb{I}_2 +
\sum_{i,i^\prime=1}^3 T_{i i^\prime} \, \sigma_i^{(1)} \otimes
\sigma_{i^\prime}^{(2)} \Big)
\end{aligned} \end{eqnarray}
where $\sigma_i$ stands for the Pauli sigma matrices. Using this
representation, the geometric measure of discord can be calculated
as:
\begin{equation} \label{Eq_GeoDiscord2}
D_{G,1:2}^{\mathrm{ZD}} = \frac{1}{4} \left( \|\vec{x}\|^2 + \|T\|^2
- k_{max} \right) ,
\end{equation}
where $\vec{x}$ equals to $(x_1, x_2, x_3)$ with $x_i = \mathrm{tr}
[\rho (\mathbb{I}_1 \otimes \sigma_i^{(2)})]$, $\|T\|^2 =
\mathrm{tr} [T^T T]$ with $T = T_{i i^\prime} \, | i \rangle \langle
i^\prime |$, $k_{max}$ is the largest eigenvalue of the matrix $K =
\vec{x} \, \vec{x}^T + T^T T$. This measure actually minimizes the
distance of a given state $\rho$ to the set of states with zero
discord (ZD) using the metric squared Hilbert-Schmidt norm, $\|\rho
- \rho^\prime\|^2 = \mathrm{tr} [\rho - \rho^\prime]^2$.

\section{Implementation of the model} \label{Model_Part2}

Almost all of the parameters of the pseudo-spin Hamiltonian
(\ref{Ham_HL}) except $J_x$ can be determined using the tools of
quantum chemistry by taking into account all the details of the
electronic structure of water ice, e.g. performing a number of
different \textit{ab initio} density functional calculations for
some of the possible $2^{12}$ proton configurations. It is also
possible to construct a realistic spectral density function based on
the molecular dynamics simulations or the density functional theory
calculations. However, this would not only increase the demand for
computational cost of our model, but also reduce its explanatory
power since the model parameters are assumed to have contributions
from the mean-field averages of the effects of the surrounding
hexamers. On the contrary, the present work aims to construct a
simple and physically insightful model with the minimum number of
parameters that can be estimated from comparisons of the predictions
of the model with experimental results. In what follows we show that
the steady state of the equation (\ref{MasterEq}) depends on only
two parameters in contrast to previous multi-parameter models and
the quantum information theoretic analysis of this state is enough
to give a quantitative description of the experimental data.

\subsection{Extension to the physical system} \label{Model2Ice}

Before elaborating on the final steady state of the equation
(\ref{MasterEq}), we first find a reasonable map between the actual
3-d structures of water ice and the present 2-d model of a single
hexamer. If we extended the model by connecting multiple hexamers in
3-d as in the water ice, we would first replace OH$^-$ ions with O
atoms in the vertices and increase the number of equally likely
locations for protons close to each vertex from two to four, i.e.,
the local constraints on a single hexamer would change. Keeping this
difference in mind, we infer the ordering dynamics in multi-hexamer
real structures from the underlying proton dynamics in single
hexamers and construct the map between the model and physical system
based on the single-hexamer proton relocation events occurring in
them.

Both of the ice I$_\text{h}$ and XI obey the ice rules. However
hexagonal rings of ice XI possess a global proton order which is
absent in ice I$_\text{h}$, i.e., H$_2$O hexamers sharing the same
3-d form and the same orientation have also the same proton
configuration. This proton order can't be preserved in the presence
of proton relocation unless each of these hexamers simultaneously
switches into another proton configuration through a collective
motion of the six protons. Thus, not only each of the 3-d hexamers,
but also the whole ice XI crystal being constituted by them is
allowed to be found only in two different configurations. The switch
between these two configurations can be mapped to the transition
between $| 01 \rangle^{\bigotimes 6}$ and $| 10 \rangle^{\bigotimes
6}$ pseudo-spin states (respectively depicted in figures
\ref{Model}-b and c) as both of them require concerted six-proton
relocation.

$| 01 \rangle^{\bigotimes 6}$ and $| 10 \rangle^{\bigotimes 6}$
pseudo-spin states span the whole subspace in which the single
hexamer satisfies the ice rules. In a sense, we assume that this ice
rule preserving subspace in the model corresponds to the
proton-ordered phase in hexagonal water ice.

Ice I$_\text{h}$ is composed of 3-d hexamers that also fulfill the
ice rules but do not show global correlation. Proton configuration
of these hexamers can be achieved from the ones in ice XI by the
proton relocation events occurring through H-bonds and keeping the
proton number in each hexamer fixed at six. In our model, similar
proton relocation events bring the states living inside the ice rule
preserving subspace into another subspace spanned by 62 pseudo-spins
representing the ionic defects. Thus, this defective pseudo-spin
subspace can be assumed to coincide with the proton-disordered phase
in hexagonal water ice.

\subsection{Asymptotic limit of the model}

Here and in the following, we consider the steady state solution of
the equation (\ref{MasterEq}). The chosen interaction with the
environment does not bring the system into a thermal equilibrium in
general. On the contrary, it divides $2^{12}$-dimensional Hilbert
space $\mathcal{H}$ into subspaces $\mathcal{H}_{\mathcal{J}}$ each
of which is independently invariant under $\{A_{j}\}_{j=1}^{12}$
operators. In the asymptotic limit, it provides a detailed balance
only inside these subspaces as below:
\begin{eqnarray} \label{roInfHex} \begin{aligned}
\rho_{\infty} &= \sum_{\mathcal{J}}
\frac{\mathrm{P}\left(\mathcal{J}\right)}{\mathcal{Z}\left(\mathcal{J}\right)}
\sum_{| \epsilon_m \rangle \in \mathcal{H}_{\mathcal{J}}} e^{- \beta
\epsilon_m} | \epsilon_m \rangle\langle \epsilon_m | ,
\end{aligned}
\end{eqnarray}
where
\begin{eqnarray} \label{roInfHexP}
\mathrm{P}\left(\mathcal{J}\right) = \sum_{| \epsilon_m \rangle \in
\mathcal{H}_{\mathcal{J}}} \langle \epsilon_m | \, \rho(t=0) |
\epsilon_m \rangle ,
\end{eqnarray}
and
\begin{eqnarray} \label{roInfHexZ}
\mathcal{Z}\left(\mathcal{J}\right) = \sum_{| \epsilon_m \rangle \in
\mathcal{H}_{\mathcal{J}}} e^{- \beta \epsilon_m} ,
\end{eqnarray}
with $\beta = 1 / k_B T$.

One of the subspaces $\mathcal{H}_{\mathcal{J}}$ consists of two
special kinds of pseudo-spin states mentioned in the previous
subsection, i.e., 2 pseudo-spin states obeying the ice rules and 62
pseudo-spin states corresponding to ionic defects. As these state
sets are assumed to map to the proton-ordered and disordered phases
of the hexagonal water ice respectively, we label their union by
$\mathcal{H}_{ice}$. If the initial state $\rho(t=0)$ lives only in
this $64$-dimensional subspace, the state of the pseudo-spins that
relax to equilibrium still stays inside the same subspace as follows
\begin{eqnarray} \label{roIce} \begin{aligned}
\rho_{\infty}^{ice} &= \sum_{| \epsilon_m \rangle \in
\mathcal{H}_{ice}} e^{- \beta \epsilon_m} | \epsilon_m
\rangle\langle \epsilon_m | \quad / \sum_{| \epsilon_m \rangle \in
\mathcal{H}_{ice}} e^{- \beta \epsilon_m} .
\end{aligned}
\end{eqnarray}

It is straightforward to show that the athermal attractor above
essentially depends on two free parameters, $J_x$ and
$J_z^{(\text{intra})}$ as all the energy eigenvalues of subspace
$\mathcal{H}_{ice}$ have a common functional dependence on the
remaining coefficients included in (\ref{Ham_HL}).

\subsection{Characterization of proton ordering/disordering} \label{Model2PhaseTrans}

A temperature dependent transition between two different phases can
be characterized by the presence of two different steady states
above and below the phase transition temperature. Although our
master equation has a unique steady state solution denoted by
$\rho_{\infty}^{ice}(T)$, it shows different features above and
below the phase transition temperature range which are reflected by
i) $P_{BF}(T)$, the probability of pseudo-spins to be found inside
the ice-rule preserving subspace and ii)
$S[\rho_{\infty}^{ice}(T)]$, the von Neumann entropy of
pseudo-spins.

In our model, each basis state represents a different configuration
of the protons. As explained in section \ref{Model2Ice}, the
2-dimensional ice rule preserving subspace spanned by $\{| 01
\rangle^{\bigotimes 6}, | 10 \rangle^{\bigotimes 6}\}$ can be mapped
to the XI phase of water ice based on the proton relocation
dynamics. Thus, the change in probability $P_{BF}(T)$ can be used as
an indicator of the proton-disordering phase transition, e.g., it
should be close to unity in XI phase and show a decrease during the
transition to I$_\text{h}$ phase.

Presence of unit probability below the phase transition temperature
range means that only $4$ particular elements of
$\rho_{\infty}^{ice}(T)$ can be nonzero. No matter how small a
deviation from unity in $P_{BF}(T)$ is, $64 \times 64 - 4 = 4092$
more elements of $\rho_{\infty}^{ice}(T)$ can take a non-zero value.
This corresponds to an enlargement in the dimension of the effective
Hilbert space from $2$ to $64$, which allows the violation of ice
rules that was mapped to the proton-disordered phase in section
\ref{Model2Ice}. Hence, there is no need to a significant decline in
$P_{BF}(T)$ within the phase transition temperature range to
indicate the proton-ordered/disordered transition, but it is
sufficient for it to gradually deviate from unity which means a
turning point behavior. When it shows a sharp decline above the
phase transition temperature range, we have effectively two
different steady states above and below the range, each of which has
different numbers of non-zero elements and lives in a different
Hilbert space.

The von Neumann entropy measures the amount of disorder,
uncertainty, or unpredictability of a generic quantum state.
Furthermore, each basis state of pseudo-spins represents a different
proton configuration in our model. Thus, $S[\rho_{\infty}^{ice}(T)]$
directly quantifies the proton-disorder.

\section{Results}

\subsection{Model parameters}

The values of the parameters used here were estimated from
comparisons of the predictions of the model with the previous
experiments carried out on the ice I$_\text{h}$/XI transition.
Actually, it isn't easy to observe this transition as protons are
expected to become classically immobile around $100 - 110$ K where a
glass transformation occurs \cite{2015_HexPhaseTransition,
1988_GlassTrans, 1997_GlassTrans}. First calorimetric measurements
\cite{1982_KOH-doped, 1984_KOH-doped} overcame this problem by using
alkali hydroxides as catalyzer and catched the transition at $72$ K.
According to recent complex dielectric constant measurements of pure
water ice reported in \cite{2015_HexPhaseTransition}, hexagonal ice
undergoes a phase transition from proton-disordered ice I$_\text{h}$
to proton-ordered ice XI at $58.9$ K, whereas a reverse
transformation occurs at $73.4$ K. Since a charge movement should
increase the imaginary part of the dielectric constant
$\varepsilon^{\prime\prime}$, these phase transitions were
determined by detecting the anomalies in the cooling and warming
curves of $\varepsilon^{\prime\prime}(T)$. Whereas a clear peak was
observed in the warming curve of $ d \varepsilon^{\prime\prime} /
dT$ at $73.4$ K, a discontinuous change occurred in the slope of
cooling curve of $ d \varepsilon^{\prime\prime} / dT$ at $58.9$ K.
The former was reflected in the warming curve of
$\varepsilon^{\prime\prime}(T)$ as a continuous increase followed by
a change in the slope, while the latter corresponded to a smooth
change in the slope of the cooling curve of
$\varepsilon^{\prime\prime}(T)$. See figure 3 in reference
\cite{2015_HexPhaseTransition} for more details.
\begin{figure}[h]
\centering
        \includegraphics[width=0.6\textwidth]{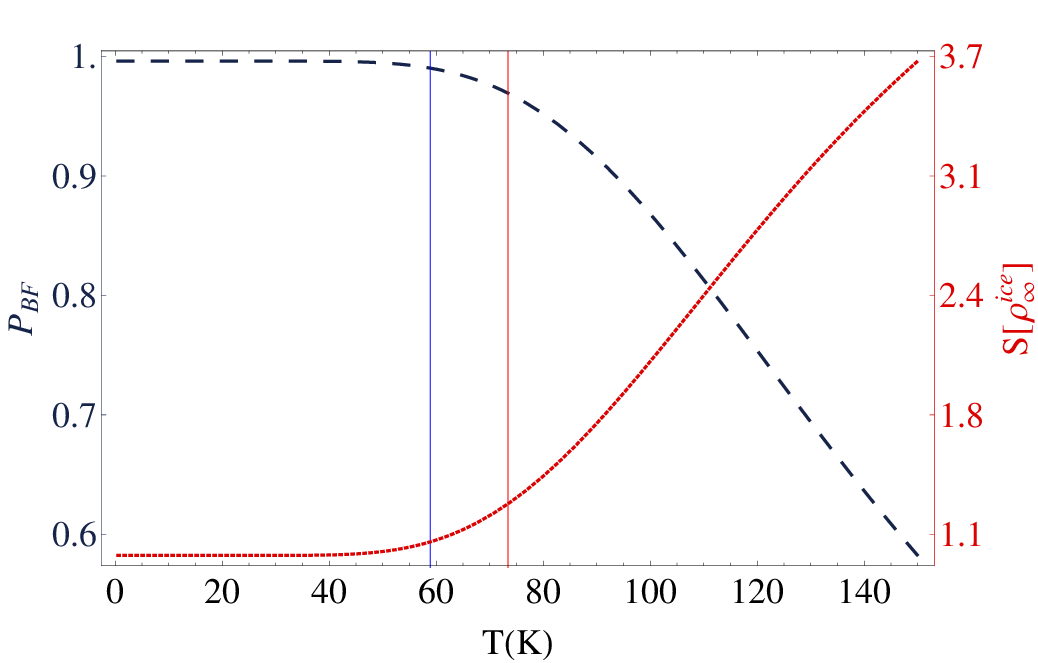}
        \caption{Estimation of the parameters based on the predictions of
        the model with experiments. XI$\rightarrow$I$_\text{h}$ and
        I$_\text{h}$$\rightarrow$XI phase transitions of hexagonal ice
        respectively occur at $58.9$ K and $73.4$ K \cite{2015_HexPhaseTransition}.
        Vertical solid lines coloured blue and red pinpoint these temperatures.
        The dashed dark blue curve is $P_{BF}(T)$, the probability of
        pseudo-spins to be found inside the ice rule preserving subspace
        spanned by $\{| 01 \rangle^{\bigotimes 6}, | 10 \rangle^{\bigotimes 6}\}$
        when they are prepared in the steady state $\rho_{\infty}^{ice}$.
        We set $J_x$ and $J_z^{(\text{intra})}$ respectively to $- 0.5$ meV and
        $+10 $ meV to observe the violation of ice rules around the
        phase transition temperatures. The dashed dark red curve shows the
        temperature dependence of the von Neumann entropy of $\rho_{\infty}^{ice}$.}
        \label{SandP}
\end{figure}

As opposed to experimental data on $\varepsilon^{\prime\prime}(T)$,
no hysteresis is expected between the warming and cooling curves of
$P_{BF}(T)$ since we focus on the steady state solution of the
master equation and do not allow the system under consideration to
be driven out of equilibrium due to quantum fluctuations. Moreover,
as our model is restricted to a single hexamer, a sharp
discontinuity in $P_{BF}(T)$ is unlikely to occur during phase
transition. Hence, instead of a first order phase transition, we
anticipate observing a smooth change in the slope of $P_{BF}(T)$
similar to that of the cooling curve of
$\varepsilon^{\prime\prime}(T)$ shown in figure 3-a in
\cite{2015_HexPhaseTransition} and in figure 2 in
\cite{2015_CorrPTInIceXI}. However, even this smooth change in our
finite size system can be treated as evidence of a
proton-disordering phase transition since it reflects the true
microscopic mechanism driving the proton-disordering process (see
sections \ref{Model2Ice} and \ref{Model2PhaseTrans} for details).
\begin{figure}[t] \centering
        \includegraphics[width=0.6\textwidth]{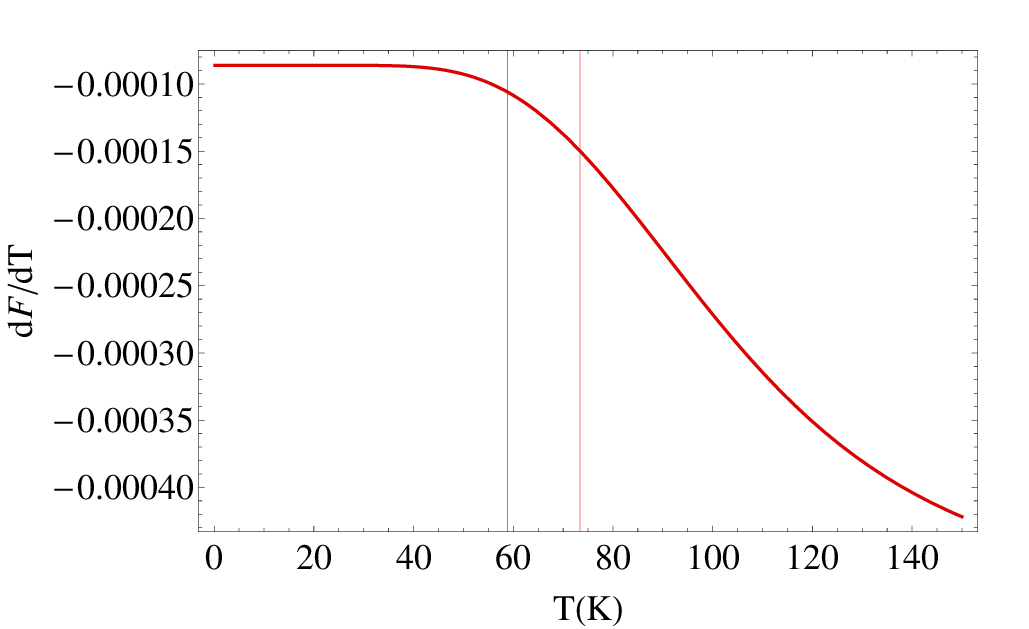}
        \caption{First derivative of the free energy with respect to
        temperature. Since we do not consider any change in the locations of
        OH$^-$ ions with temperature, we do not include a pressure-volume term
        in the free energy, i.e., $F = U - k_B T S$, where $S$ is the von Neumann
        entropy.}
        \label{F}
\end{figure}

In this respect, we fixed our free parameters $J_x$ and
$J_z^{(\text{intra})}$ respectively to $- 0.5$ meV and $+ 10$ meV to
reproduce the expected trend of $P_{BF}(T)$ using the steady state
$\rho_{\infty}^{ice}$ as shown in figure \ref{SandP}. The state of
pseudo-spins stays inside the ice rule preserving subspace with a
unit probability while $T < 58.9$ K. The states outside this
subspace violate the ice rules and gradually become available
between the blue and red lines. Above the temperature of $\approx
73.4$, $P_{BF}(T)$ shows a sharp decline that corresponds to an
ever-increasing population of ice rule violating states. Details of
the procedure used in this parameter estimation are given in section
\ref{II-A} in the Electronic Supplementary Material (ESM).

The behaviour of the von Neumann entropy of $\rho_{\infty}^{ice}(T)$
supports the arguments above for the fixed values of $J_x$ and
$J_z^{(\text{intra})}$. According to figure \ref{SandP}, it remains
at unity until $58.9$ K. Note that the pseudo-spins live inside the
subspace spanned by $\{| 01 \rangle^{\bigotimes 6}, | 10
\rangle^{\bigotimes 6}\}$ in the same temperature range. Then this
unit disorder is possible only if the pseudo-spins can exist in only
two orthogonal states living inside this subspace with an equal
probability, i.e., the configurations given in figures \ref{Model}-b
and c or two of their coherent superpositions orthogonal to each
other are equally likely for the protons.
$S[\rho_{\infty}^{ice}(T)]$ rises slowly with further increases of
temperature until $73.4$ K. Hence, a smooth change occurs in the
proton-disorder around $58.9 - 73.4$ K. After that, the slope of
$S[\rho_{\infty}^{ice}(T)]$ is approximately constant, which shows a
rapid increase in the proton disorder. Note that the first
derivative of free energy of $\rho_{\infty}^{ice}(T)$ has the same
temperature dependence with $P_{BF}(T)$ and
$S[\rho_{\infty}^{ice}(T)]$ as shown in figure \ref{F}.
\begin{figure}[t] \centering
        \includegraphics[width=\textwidth]{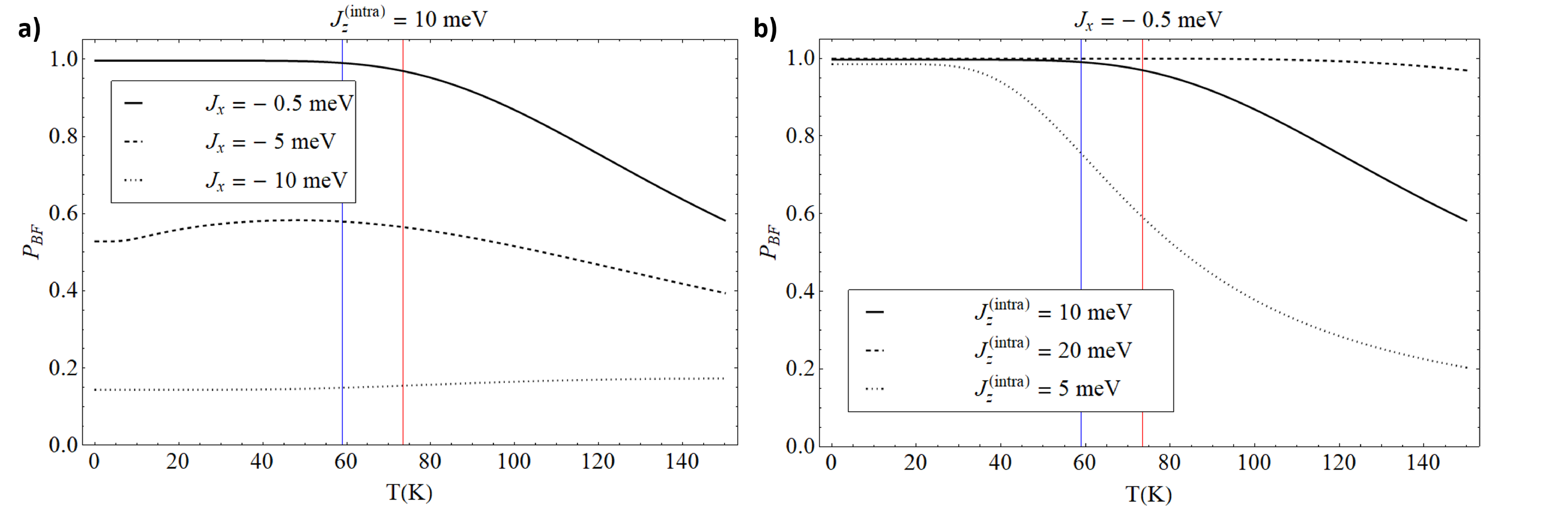}
        \caption{Sensitivity of the behaviour of probability $P_{BF}(T)$
        to the fixed values of free parameters $J_x$ (a) and
        $J_z^{(\text{intra})}$ (b).}
        \label{Pv2}
\end{figure}

Estimation of the free parameters may still look arbitrary at first
glance. However, if we decrease the value of $J_x$, $P_{BF}(T)$
cannot get close to unity at any temperature (figure \ref{Pv2}-a),
which means that the ice rules are always violated and the system
can never end up in XI phase. On the contrary, a change in the value
of $J_z^{(\text{intra})}$ sets the temperature at which the ice
rules begin to be violated, apart from the experimentally determined
phase transition temperatures (figure \ref{Pv2}-b). Hence, the
expected temperature dependence of $P_{BF}$ exhibits a sensitivity
to our free parameters, i.e., deviations from the fixed values of
either $J_x$ or $J_z^{(\text{intra})}$ that are much smaller than
the energy of a H-bond rule out any prediction, preventing the
appearance of a slow decline in $P_{BF}(T)$ from unity around $58.9
- 73.4$ K. Also, this behaviour cannot reappear when the second
parameter is also allowed to deviate from its fixed value at the
same time. Please see section \ref{II-B} in the Electronic
Supplementary Material (ESM) for the details.

Note that the values of $J_x = - 0.5$ meV and $J_z^{(\text{intra})}
= + 10$ meV are fixed in this way, and do not only stand for the
bare coefficients of a single hexamer but also have contributions
from the mean-field averages of the effects of the surrounding
hexamers.

\subsection{Quantum aspects of the proton mobility}

We summarized several theoretical and experimental findings
suggesting the likelihood of proton tunneling in hexagonal water ice
in section \ref{Intro}. Here, we address two of them using the tools
of quantum information theory described in section \ref{QIT}.

The first claim concerns the role of proton tunneling in
I$_\text{h}$/XI phase transition. In reference
\cite{2006_PTInIceIh}, the motion of the protons was first mapped
into a pseudo-spin model, as we do, but then converted to a gauge
theory problem. This description allowed the authors to characterize
the ordered and disordered phases respectively by confined and
deconfined behaviours of ionic defects in the ground state of the
system. It was then found that the phase transition under
consideration is possible only if the protons tunnel through H-bonds
with a rate greater than a critical value (see figure 6-b in
\cite{2006_PTInIceIh}). Since the protons are expected to become
classically immobile around $100 - 110$ K where a glass
transformation occurs \cite{2015_HexPhaseTransition,
1988_GlassTrans, 1997_GlassTrans}, this is a reasonable claim. Our
predictions shown in figure \ref{Pv2}-a suggest that it may be also
unlikely for the hexagonal water ice to end up in XI phase unless
the tunneling rate is less than another critical value. Thus,
further investigation of the quantum aspects of the proton mobility
in I$_\text{h}$/XI transition using quantum information theory may
yield new knowledge about this topic.
\begin{figure}[t] \centering
        \includegraphics[width=0.64\textwidth]{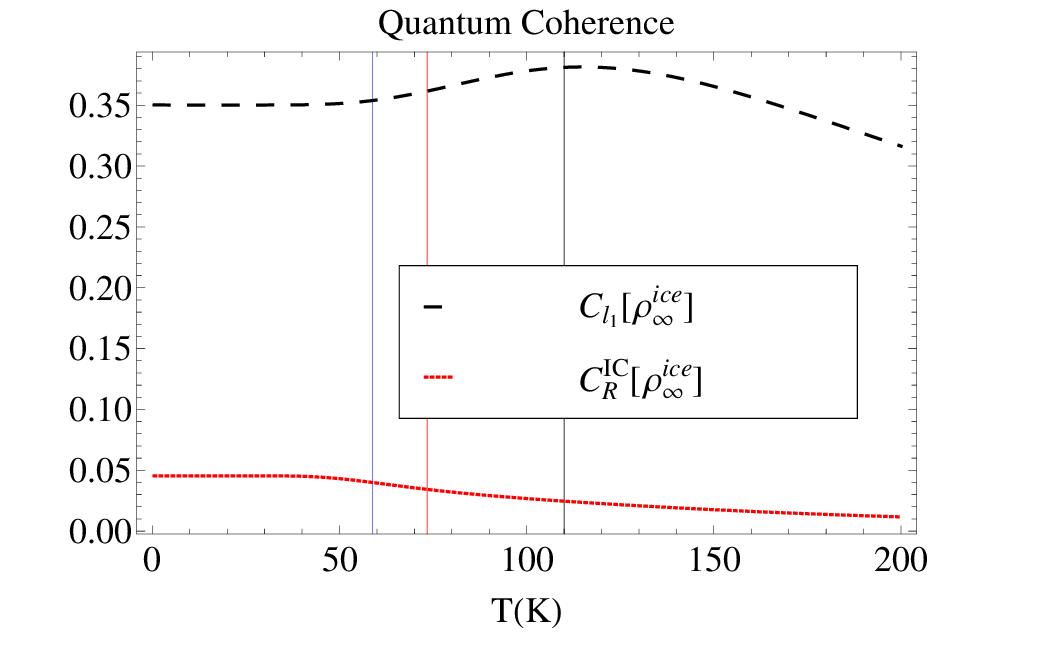}
        \caption{Coherent quantum effects on proton mobility in a
        H$_2$O hexamer. Vertical solid lines coloured blue and
        red indicate the phase transition temperatures \cite{2015_HexPhaseTransition},
        whereas the vertical black line is anchored to the temperature
        of the glass transition at which proton mobility is expected
        to diminish \cite{2015_HexPhaseTransition, 1988_GlassTrans, 1997_GlassTrans}.}
        \label{Coh}
\end{figure}

The second claim that will be addressed here is the presence of
concerted six-proton tunneling at low temperatures in XI phase.
Dielectric constant measurements that determined the phase
transition temperatures in pure water ice
\cite{2015_HexPhaseTransition} were extended down to $5$ K in
reference \cite{2015_CorrPTInIceXI}, and an anomaly was observed in
the cooling and warming curves of $\varepsilon^{\prime\prime}(T)$ in
the form of a minimum around $20$ K. The monotonic behaviour of the
real part of the dielectric constant observed in the same data and
disappearance of the anomaly in the repeat measurements on heavy ice
were explained by the back and forth tunneling of protons in groups
of six. As mentioned before, the unit disorder of
$\rho_{\infty}^{ice}$ at low temperatures (figure \ref{SandP}) may
indicate the presence of two equally likely superpositions of the
configurations given in figures \ref{Model}-b and c. Also, each such
superposition represents the correlated tunneling of six protons,
and quantum information theory is able to study the nature and
extent of this correlation (see section \ref{QIT}).

To provide a first insight into the quantum aspects of proton
mobility in a hexameric H$_2$O loop, we apply the $l_1$ norm and
relative entropy of coherence on $\rho_{\infty}^{ice}$ as shown in
figure \ref{Coh}. $C_{R}^{\mathrm{IC}}\![\rho_{\infty}^{ice}]$,
which quantifies the distinguishability of $\rho_{\infty}^{ice}$
from its completely decohered version, remains constant throughout
the XI phase and steadily decreases with increasing temperature.
Hence, there is a rise in the loss of collective quantumness in the
global proton mobility starting with the transition from XI phase to
I$_\text{h}$ phase which lasts thereafter. Conversely,
$C_{l_1}\![\rho_{\infty}^{ice}]$, which is the sum of quantum
coherences in individual transitions between proton configuration
pairs, shows a different behaviour with respect to temperature. It
increases with the XI$\rightarrow$I$_\text{h}$ phase transition and
reaches a peak around glass transition. Actually, this behaviour
seems to be consistent with the experimental data related to real
proton mobility, which indicates a local maximum between $60 - 110$
K (see figure 3-a in \cite{2015_HexPhaseTransition} and figure 2 in
\cite{2015_CorrPTInIceXI}) at where the protons are expected to be
classically immobile.

The deviation of the temperature dependence of
$C_{l_1}\![\rho_{\infty}^{ice}]$ from the experimental data below
$60$ K is related to the finite size of our model. Although the
increase in $\varepsilon^{\prime\prime}(T)$ during the cooling from
$20$ K to $5$ K arises from the increasing number of protons
involved in the correlated six-proton tunneling events
\cite{2015_CorrPTInIceXI}, our results are limited to a single
hexamer including only six protons. The loss of similarity between
the curves of $C_{l_1}\![\rho_{\infty}^{ice}(T)]$ and
$\varepsilon^{\prime\prime}(T)$ above $110$ K also originates from
the restrictions on our model. The rise in
$\varepsilon^{\prime\prime}(T)$ after the glass transition
\cite{2015_CorrPTInIceXI, 2015_HexPhaseTransition} is likely stem
from thermally activated proton hopping, which is expected to
suppress quantum coherent proton mobility at these temperatures but
outside our scope. Note that the classical motion of the protons
enters into our model in the form of incoherent superpositions of
pseudo-spins that are generated from coherent superpositions as a
result of decoherence.
\begin{figure}[t] \centering
        \includegraphics[width=0.63\textwidth]{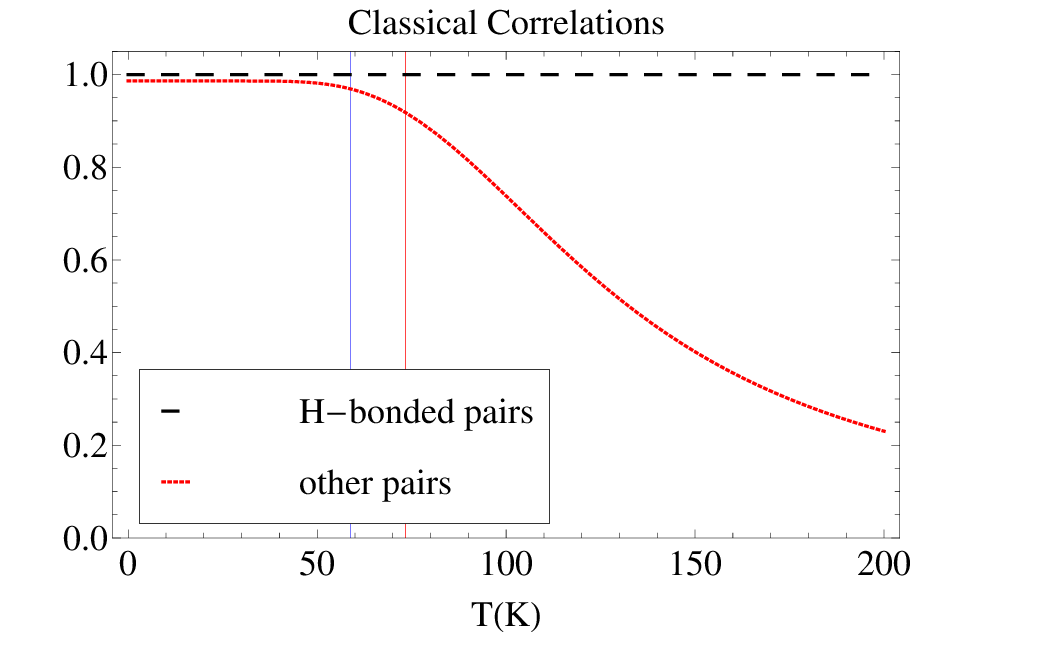}
        \caption{Pairwise classical correlations. The dashed black
        curve corresponds to the correlations in a single
        H-bond. Correlations between two protons, each of which
        belongs to a different H-bond, is shown by the dashed dark red
        curve.}
        \label{CC}
\end{figure}
\begin{figure}[t] \centering
        \includegraphics[width=0.63\textwidth]{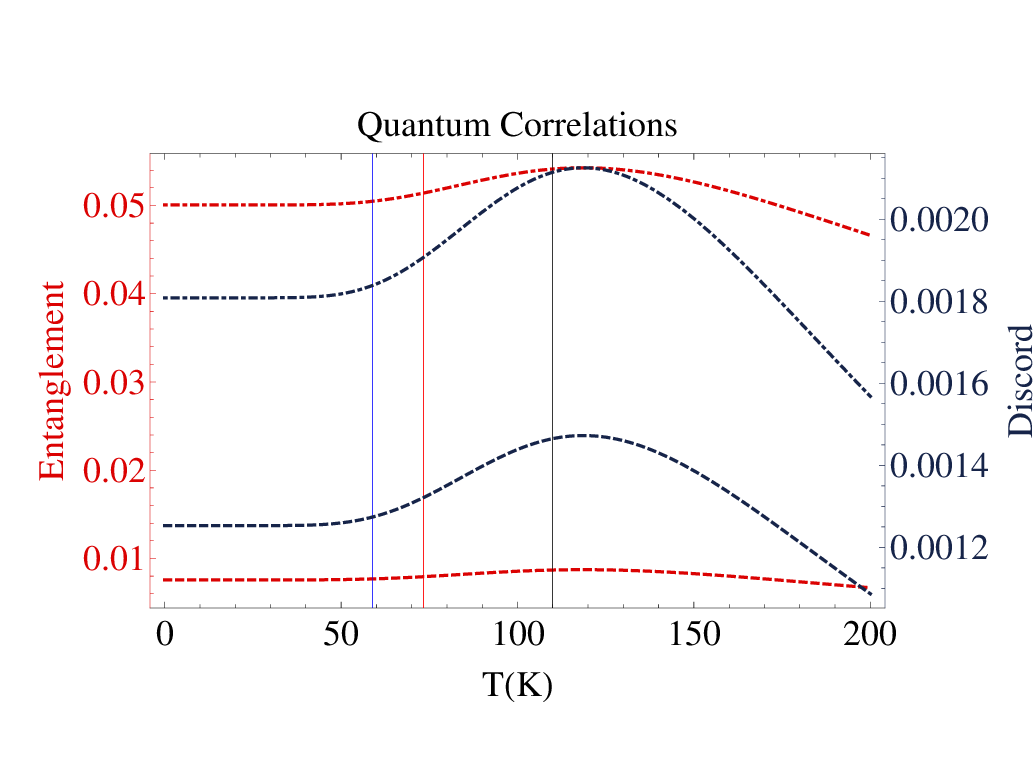}
        \caption{Pairwise quantum correlations in a single H-bond. There is no quantum
        correlation found between proton pairs belonging to
        different H-bonds. The dot-dashed dark red curve is the concurrence between pseudo-spins
        lying on the same edge, while the dashed red curve is the
        entanglement of formation between them. Their quantum
        correlations beyond entanglement are measured by quantum
        discord and its geometric measure, which are displayed respectively
        as dot-dashed and dashed dark blue curves.}
        \label{QC}
\end{figure}

On the other hand, the uptick in $C_{l_1}\![\rho_{\infty}^{ice}]$
between the blue and red solid lines in figure \ref{Coh} offers
fresh insights about the importance of coherent proton mobility on
the proton ordering dynamics \cite{2006_PTInIceIh}, i.e., although
the tunneling coefficient is fixed initially, there is an increase
in the amount of coherence generated by tunneling events during the
phase transition from proton-disordered ice I$_\text{h}$ to
proton-ordered ice XI. Also, pinning of
$C_{l_1}\![\rho_{\infty}^{ice}(T)]$ at a nonzero value below $60$ K
eliminates the possibility that the pseudo-spins exist in a maximal
mixture of the basis states $| 01 \rangle^{\bigotimes 6}$ and $| 10
\rangle^{\bigotimes 6}$. Hence, the presence of unit disorder below
$60$ K in figure \ref{Pv2}-a should come from a maximal mixture of
two orthogonal superpositions of these basis states, i.e.,
$\rho_{\infty}^{ice} = 1/2 | \psi_+ \rangle \langle \psi_+ | + 1/2 |
\psi_- \rangle \langle \psi_- |$ where $| \psi_\pm \rangle$ are two
coherent superpositions such as $\left(| 01 \rangle^{\bigotimes 6}
\pm | 10 \rangle^{\bigotimes 6}\right) / \sqrt{2}$. Note that each
of the superposition states involved in this mixture can be
interpreted as a concerted tunneling of six protons back and forth
between the configurations represented by the states $| 01
\rangle^{\bigotimes 6}$ and $| 10 \rangle^{\bigotimes 6}$. This
observation is in accordance with the dielectric anomaly measured in
the form of a minimum around $20$ K \cite{2015_CorrPTInIceXI} where
we suspect a concerted quantum tunneling of six protons could be
occurring in each hexamer.

A deeper understanding of the overall behaviour of
$C_{l_1}\![\rho_{\infty}^{ice}(T)]$ requires investigation of the
classical and quantum correlations between pseudo-spin pairs
respectively shown in figures \ref{CC} and -\ref{QC}. Only classical
correlations appear between the pseudo-spins lying on different
edges according to these figures. This is actually what we expect to
see as we prevent the formation of quantum correlations between
these pseudo-spin pairs by setting the Hamiltonian parameter
$J_{j,j+1}^{(\text{vertex})}$ to zero and taking the O$-$H stretch
vibrations as independent from each other. What is unexpected about
these results is that the probability change observed in figure
\ref{SandP} resembles the temperature-dependent behaviour of the
classical correlations in figure \ref{CC}, where the mutual
information $J$ of the reduced state of the corresponding
pseudo-spins is constant around unity throughout the XI phase and
starts falling down during the XI/I$_\text{h}$ transition. This
means that there is an approximately maximal amount of classical
correlations between the motions of two different protons that
belong to different H-bonds below $60$ K.

Beside this, classical correlations between edge sharing
pseudo-spins seem to be fixed at unity independent of the
temperature. Hence, classical correlations generated by the proton
motion in an individual H-bond are invariant under any change in
temperature. On the other hand, regardless of the measure that is
used to quantify quantum correlations between edge sharing
pseudo-spin pairs, quantum correlations generated by proton
tunneling through individual H-bonds are found to be quite low.
However, each curve in figure \ref{QC} shows a similarity with
coherent proton dynamics described by
$C_{l_1}\![\rho_{\infty}^{ice}(T)]$ in figure \ref{Coh}. Thus,
although the quantum correlations in H-bonds seem to be
insignificant when compared to their classical counterparts,
temperature dependence of quantum coherent proton mobility still
originates from them.

Based on these observations, we surmise that all the individual
proton tunnelings throughout six H-bonds found in a single hexamer
become classically correlated at low temperatures. These
correlations start to weaken during the phase transition from
proton-ordered phase ice XI to proton-disordered phase ice
I$_\text{h}$. At the same time, quantum correlations between
H-bonded atoms become stronger, reaching a maximum during the glass
transition around $110 - 120$ K.

\section{Future Directions}

The effect of O$-$H stretch vibrations is incorporated into the
model in the form of Holstein-type local phonon-proton couplings.
Such local interactions can originate from the second quantization
of the small site displacements after expanding on-site energies
$W_j$ around some reference set of coordinates. However, transfer
integrals $J_{j,j+1}^{(\text{edge})}$ are also likely to be
perturbed by both O$-$H and O$-$O vibrations. Moreover, these
perturbations should have a different kind of nature, which can be
described by non-local phonon-proton couplings well-known as
Peierls-type interaction. Unlike the Holstein-type interaction, this
interaction isn't necessarily destructive and can facilitate proton
tunneling between sites. In fact, the fluctuations of the O$-$H and
O$-$O bond lengths are expected to have a significant effect on the
proton dynamics in real water ice structures. A natural direction to
pursue future work is a new quantum master equation approach to open
system dynamics of the protons in the presence of both local and
non-local phonon couplings. Such a mixed Holstein-Peierls model will
reveal the competition between the local and non-local phonon
couplings that may be key to understanding this system.

Our first attempts to develop a mixed Holstein-Peierls model using a
two qubit model system given in the Electronic Supplementary
Material (ESM) show that although nonlocal couplings change the
dynamics of the system, the static structure of the final steady
state remains same (see section \ref{Sec_OpenHBModel}). Hence, the
bond fluctuations appear not to affect the predictions of the
present model unless we do not move on to probe the true proton
dynamics around the hexagonal ring in real-time.

The simple pseudo-spin model approach can be readily employed
together with the quantum chemical techniques treating electrons and
protons quantum mechanically. This additional technique will allow
us to give a realistic estimate of both the parameters of the
self-Hamiltonian and the form of the spectral density function. It
will then be possible to probe the true proton dynamics around the
hexagonal ring in real-time. This offers a new opportunity that
previous studies didn't yield. However, our assumptions about the
Markovianity of the open system dynamics may not be justified in
this case as we will not work in the long-time regime. Thus, this
direction also includes description of the proton dynamics using a
non-Markovian evolution.

Besides this, our one-qubit pseudo-spin representation of proton
locations is suitable to extend the present model to include nuclear
spin degrees of freedom, which are usually ignored by the current
modelling approaches in the water literature. Actually, the
experimentally determined ratio of ortho/para spin states of
isolated single water molecules exhibits a temperature dependence
which shares similarities with the proton mobility under
consideration in this study. In this respect, we are planning to
extend the present model to investigate the possible effects of
nuclear spins on proton-ordering dynamics in water ice.

\section{Conclusion}

We constructed a simple pseudo-spin model to investigate both the
possibility and the nature of concerted six-proton tunneling in a
hexameric H$_2$O employing the tools of quantum information theory.
We demonstrated that the static structure of the final steady state
of the chosen master equation depends on only two parameters and the
quantum information theoretic analysis of this state is sufficient
to give a quantitative description of experimental data.

The role of the external environment on the concerted six-proton
tunneling was clearly unveiled as this tunneling process was not
imposed by the self-Hamiltonian but emerged naturally in the
long-time limit of the low-temperature dynamics of the open system.
Thus, phonon-assistance was found to be central in driving the
concerted proton tunneling up to the temperature of the phase
transition from ice XI to ice I$_\text{h}$. Moreover, it was found
to be associated with the emergence of ice rules governing the
arrangement of atoms in water ice.

Remaining within the framework of the pseudo-spin model enabled us
to approach the correlation problem using the tools of quantum
information theory. In turn, we inferred that the $l_1$ norm of
coherence \cite{Plenio-2014} is sufficient to capture the behaviour
of coherent proton mobility observed in experiments
\cite{2015_HexPhaseTransition, 2015_CorrPTInIceXI}. We also
discriminated between quantum and classical correlations in
concerted proton tunneling. It was found that the correlations
between six proton tunneling events are not inherently quantum in
character. Instead, individual tunneling events were allowed to be
classically correlated only. Low rates and strong correlations were
observed for quantum tunneling events below a critical temperature
corresponding to phase transition. Beyond this critical temperature,
simulations showed a weakening in the correlations, but an increase
in rates. Overall this induces a total increment in the coherent
proton mobility until achieving a full proton disorder.

The finding sheds light on the nature of correlations between the
individual tunneling events that can't be addressed by previous
studies. This also paves the way to investigating the proton
dynamics in real-time to advance our understanding of many-proton
tunneling in water ice. \\

\begin{acknowledgments}

O.P. thanks Alptekin Y{\i}ld{\i}z for many insightful discussions
about the dielectric constant measurements and thanks TUBITAK
2214-Program for financial support. T.F. and V.V. thank the Oxford
Martin Programme on Bio-Inspired Quantum Technologies, the EPSRC and
the Singapore Ministry of Education and National Research Foundation
for financial support.

\end{acknowledgments}

\pagebreak

\begin{center}
  \textbf{\large Electronic Supplementary Material}\\[.2cm]
  available online at https://dx.doi.org/10.6084/m9.figshare.c.4487114 \\[.1cm]
\end{center}

\setcounter{equation}{0}
\setcounter{figure}{0}
\setcounter{table}{0}
\setcounter{page}{1}
\renewcommand{\theequation}{S\arabic{equation}}
\renewcommand{\thefigure}{S\arabic{figure}}
\renewcommand{\bibnumfmt}[1]{[S#1]}

\section{Open System Dynamics of Proton Motion} \label{Sec_OpenHBModel}

The state of pseudo-spins lives in a $2^{12}$-dimensional Hilbert
space $\mathcal{H}$, that is to say that the density matrix $\rho$
describing this state has $2^{12} \times 2^{12}$ elements. In this
respect, it is not straightforward to present the details of its
open system dynamics. For the sake of simplicity, and without loss
of generality, we will focus on the motion of a single proton
between the locations 1 and 2 in the hexamer in what follows (see
Fig. \ref{Model}-a for the details). Hence, instead of working with
the twelve-site self-Hamiltonian $H_{H\!ex}$ given in Eq.
(\ref{Ham_HL0}), we will use the following two-site Hamiltonian
\begin{eqnarray}\label{Eq_Ham_HB1} \begin{aligned}
H_{H\!B} &= \sum_{j=1}^{2} W_j \, \hat{n}_j - J_{12} (a_1^{\dagger}
a_{2} + a_1 a_{2}^{\dagger}) + V_{12} \, \hat{n}_1 \hat{n}_{2} +
\lambda \, \mathbb{I}_{12}
\end{aligned}
\end{eqnarray}
to describe the closed system dynamics. Note that asymmetric version
of this Hamiltonian ($W_1 \neq W_2$) was also used in
\cite{2018_Pusuluk} to investigate the role of proton tunneling in
biological catalysis.

After applying the Jordan-Wigner transformation given in Eq.
(\ref{anticomm}) on this two-site Hamiltonian, we end up with the
two-qubit Hamiltonian
\begin{eqnarray} \begin{aligned} \label{Eq_Ham_HB2}
H_{H\!B} = J_x \left( \sigma^{(1)}_{x} \otimes \sigma^{(2)}_{x} +
\sigma^{(1)}_{y} \otimes \sigma^{(2)}_{y} \right) + J_z \,
\sigma^{(1)}_{z} \otimes \sigma^{(2)}_{z} + B \left(
\sigma^{(1)}_{z} + \sigma^{(2)}_{z} \right) + \tilde{\lambda} ,
\end{aligned}
\end{eqnarray}
where $J_x = J_{12} / 2$, $J_z = V_{12} / 4$, $B = - (2 W + V_{12})
/ 4$, and $\tilde{\lambda} = \lambda + (4 W + V_{12}) / 4$.
Eigensystem of this Hamiltonian can be written in an increasing
order of the eigenvalues as
\begin{eqnarray} \label{Eq_Eig_Ham_HB2}
\begin{cases}
e_1 = - J_z + 2 J_{x} + \tilde{\lambda}, & \mbox{} |e_1\rangle
\mbox{} = (|01\rangle + |10\rangle) / \sqrt{2} , \\
e_2 = - J_z - 2 J_{x} + \tilde{\lambda}, & \mbox{} |e_2\rangle
\mbox{} = (|01\rangle - |10\rangle) / \sqrt{2} , \\
e_3 = - 2 B + J_z + \tilde{\lambda}, & \mbox{} |e_3\rangle \mbox{} =
|11\rangle , \\
e_4 = + 2 B + J_z + \tilde{\lambda}, & \mbox{} |e_4\rangle \mbox{} =
|00\rangle .
\end{cases}
\end{eqnarray}

\subsection{Local proton-phonon coupling} \label{Sec_OpenHBModel_I}

First, we examine the O$-$H stretch vibrations by considering them
as two independent thermal baths existing around the proton
locations and having the individual self-Hamiltonians $H_B^{(j)}$
given in Eq. (\ref{HamB}). Also, we will describe the interaction of
the proton with these vibrations using the interaction Hamiltonian
given in Eq. (\ref{HamI1}):
\begin{eqnarray}\begin{aligned} \label{Eq_HamI1}
H^{local}_{I} = \sum_{j} \hat{n}_{j} \,\,\, \sum_k  \, \left(g_{j,
k} b_{j, k}^{\dagger} + g^*_{j, k} b_{j, k}\right) \propto \sum_{j}
\sigma_z^{(j)} \sum_k  \, \left(g_{j, k} b_{j, k}^{\dagger} +
g^*_{j, k} b_{j, k}\right) .
\end{aligned}
\end{eqnarray}

\subsubsection{Bath operators in interaction picture}

If we switch into the interaction picture, the bath operators $B_j =
\sum_k \, (g_{j, k} b_{j, k}^{\dagger} + g^*_{j, k} b_{j, k})$
become
\begin{eqnarray} \begin{aligned} \label{Eq_Bt1}
B_{j}(t) &= e^{X} B_{j} e^{- X} \\
&= B_{j} + [X, B_{j}] + [X, [X, B_{j}]]/2! + [X, [X, [X, B_{j}]]]/3!
+ ... \end{aligned}
\end{eqnarray}
where $X = i \, H_\mathcal{B}^{(j)} \, t / \hbar$. To evaluate the
commutators above, first we need to find the commutators
$[H_\mathcal{B}^{(j)}, b_{j,k^\prime}^{\dagger}]$ and
$[H_\mathcal{B}^{(j)}, b_{j,k^\prime}]$. In this respect, the
bosonic commutation relations imply that
\begin{eqnarray} \begin{aligned} \label{Eq_CommHAndbd}
[H_\mathcal{B}^{(j)}, b_{j,k^\prime}^{\dagger}] &= \sum_k \hbar
\omega_{j,k}
\, [b_{j,k}^{\dagger} b_{j,k}, \, b_{j,k^\prime}^{\dagger}] \\
&= \sum_k \hbar \omega_{j,k} \, \left(b_{j,k}^{\dagger} [b_{j,k}, \,
b_{j,k^\prime}^{\dagger}] + [b_{j,k}^{\dagger}, \,
b_{j,k^\prime}^{\dagger}] b_{j,k} \right)
\\
&= \sum_k \hbar \omega_{j,k} \, b_{j,k}^{\dagger}
\delta_{k,k^\prime} = + \hbar \omega_{j,k^\prime} \,
b_{j,k^\prime}^{\dagger} ,
\end{aligned}
\end{eqnarray}
\begin{eqnarray} \begin{aligned} \label{Eq_CommHAndb}
[H_\mathcal{B}^{(j)}, b_{j,k^\prime}] &= \sum_k \hbar \omega_{j,k}
\, [b_{j,k}^{\dagger} b_{j,k}, \, b_{j,k^\prime}] \\
&= \sum_k \hbar \omega_{j,k} \, \left(b_{j,k}^{\dagger} [b_{j,k}, \,
b_{j,k^\prime}] + [b_{j,k}^{\dagger}, \, b_{j,k^\prime}] b_{j,k}
\right)
\\
&= \sum_k \hbar \omega_{j,k} \, \left( -\delta_{k,k^\prime}\right)
b_{j,k} = - \hbar \omega_{j,k^\prime} \, b_{j,k^\prime} .
\end{aligned}
\end{eqnarray}

Then, it is easy to calculate the commutators in (\ref{Eq_Bt1})
after writing them in terms of (\ref{Eq_CommHAndbd}) and
(\ref{Eq_CommHAndb}):
\begin{eqnarray} \begin{aligned} \label{Eq_CommXAndB}
[X, B_{j}] &= (i \, t / \hbar) [H_\mathcal{B}^{(j)}, B_{j}] \\
&= \sum_{k} (i \, t / \hbar) \left(g^*_{j,k} [H_\mathcal{B}^{(j)},
b_{j,k}] + g_{j,k} [H_\mathcal{B}^{(j)}, b_{j,k}^{\dagger}]\right)
\\ &= \sum_{k} (i \, \omega_{j,k} \, t) \left(- g^*_{j,k} b_{j,k} +
g_{j,k} b_{j,k}^{\dagger}\right) , \end{aligned}
\end{eqnarray}
\begin{eqnarray} \begin{aligned} \label{Eq_CommXAndB2}
[X, [X, B_{j}]] &= (i \, t / \hbar) [H_\mathcal{B}^{(j)}, [X, B_{j}]] \\
&= \sum_{k} (i^2 \, \omega_{j,k} \, t^2 / \hbar) \left(- g^*_{j,k}
[H_\mathcal{B}^{(j)}, b_{j,k}] +
g_{j,k} [H_\mathcal{B}^{(j)}, b_{j,k}^{\dagger}]\right) \\
&= \sum_{k} (i^2 \, \omega_{j,k}^2 \, t^2) \left(+ g^*_{j,k} b_{j,k}
+ g_{j,k} b_{j,k}^{\dagger}\right) ,
\end{aligned}
\end{eqnarray}
\begin{eqnarray} \begin{aligned} \label{Eq_CommXAndB3}
[X, [X, [X, B_{j}]]] &= (i \, t / \hbar) [H_\mathcal{B}^{(j)}, [X, [X, B_{j}]]] \\
&= \sum_{k} (i^3 \, \omega_{j,k}^2 \, t^3 / \hbar) \left(+ g^*_{j,k}
[H_\mathcal{B}^{(j)}, b_{j,k}] + g_{j,k} [H_\mathcal{B}^{(j)},
b_{j,k}^{\dagger}]\right) \\
&= \sum_{k} (i^3 \, \omega_{j,k}^3 \, t^3) \left(- g^*_{j,k} b_{j,k}
+ g_{j,k} b_{j,k}^{\dagger}\right) .
\end{aligned}
\end{eqnarray}

By substituting these commutators into (\ref{Eq_Bt1}) and collecting
terms involving $b_{j,k}$ and $b_{j,k}^{\dagger}$ together, we end
up with the interaction picture operators given by
\begin{eqnarray} \begin{aligned} \label{Eq_Bt2}
B_{j}(t) = \sum_{k} \sum_{l=0}^{\infty} \frac{(- i \, \omega_{j,k}
\, t)^l}{l!} g^*_{j,k}  b_{j,k} + \sum_{k} \sum_{l=0}^{\infty}
\frac{(+ i \, \omega_{j,k} \, t)^l}{l!} g_{j,k} b_{j,k}^{\dagger} =
\sum_k g^*_{j,k} e^{- i \, \omega_{j,k} \, t} b_{j,k} + g_{j,k} e^{+
i \, \omega_{j,k} \, t} b_{j,k}^{\dagger} .
\end{aligned} \end{eqnarray}

\subsubsection{Thermal bath correlation function and dissipation
rates}

To calculate the bath correlation function $\langle
B_{j}^{\dagger}(t) B_{j^\prime}(0) \rangle_{\text{th}}$, we will use
the following thermal expectations:
\begin{eqnarray} \begin{aligned} \label{Eq_ThExp}
\langle b_{j,k} b_{j,k} \rangle_{\text{th}} &= 0 , & \langle
b_{j,k}^{\dagger} b_{j,k} \rangle_{\text{th}} &=
N_{j}(\omega_{j,k}) , \\
\langle b_{j,k} b_{j,k}^{\dagger} \rangle_{\text{th}} &= 1 +
N_{j}(\omega_{j,k}) , & \langle b_{j,k}^{\dagger} b_{j,k}^{\dagger}
\rangle_{\text{th}} &= 0 ,
\end{aligned}\end{eqnarray}
where $N_{j}(\omega_{j,k})$ is the average number of phonons with
energy $\hbar \omega_{j,k}$ for the Bose-Einstein statistics and
equals to $1 / (e^{\beta \hbar \omega_{j,k}} - 1)$. Then, for
independent baths, the bath correlation function becomes:
\begin{align}
\langle B_{j}^{\dagger}(t) B_{j^\prime}(0) \rangle_{\text{th}} &=
\sum_{k,k^\prime} \left\langle \big(g^*_{j,k} e^{- i \, \omega_{j,k}
\, t} b_{j,k} + g_{j,k} e^{+ i \, \omega_{j,k} \, t}
b_{j,k}^{\dagger} \big) \big(g_{j^\prime,k^\prime}
b_{j^\prime,k^\prime}^{\dagger} + g^*_{j^\prime,k^\prime}
b_{j^\prime,k^\prime} \big)
\right\rangle_{\text{th}} \nonumber \\
&= \sum_{k,k^\prime} \big( g_{j,k} g^*_{j^\prime,k^\prime} e^{+ i \,
\omega_{j,k} \, t} \langle b_{j,k}^{\dagger}, b_{j^\prime,k^\prime}
\rangle_{\text{th}} + g^*_{j,k} g_{j^\prime,k^\prime} e^{- i \,
\omega_{j,k} \, t} \langle b_{j,k}, b_{j^\prime,k^\prime}^{\dagger}
\rangle_{\text{th}} \nonumber \\
&\quad \;\;\;\; + g_{j,k} g_{j^\prime,k^\prime} e^{+ i \,
\omega_{j,k} \, t} \langle b_{j,k}^{\dagger},
b_{j^\prime,k^\prime}^{\dagger} \rangle_{\text{th}} + g^*_{j,k}
g_{j^\prime,k^\prime}^* e^{- i \, \omega_{j,k} \, t}
\langle b_{j,k}, b_{j^\prime,k^\prime} \rangle_{\text{th}} \big) \nonumber \\
&= \sum_k |g_{j,k}|^2 \Big(e^{- i \, \omega_{j,k} \, t} \big(1 +
N_{j}(\omega_{j,k})\big) + e^{+ i \, \omega_{j,k} \, t}
N_{j}(\omega_{j,k})\Big) \delta_{j j^\prime} . \label{Eq_BCorr2}
\end{align}

Dissipation rates $\gamma_{j j^\prime}$, half of the real part of
one-sided Fourier transforms of $\langle B_{j}^{\dagger}(t)
B_{j^\prime}(0) \rangle_{\text{th}}$, can be calculated by using
(\ref{Eq_BCorr2}) as
\begin{align} \label{Eq_Rates2}
\gamma_{j j^\prime}(\omega) &= \Gamma_{j j^{\prime}}(\omega) +
\Gamma^*_{j^{\prime} \! j}(\omega) \nonumber \\
&= \frac{1}{\hbar^2} \int_{- \infty}^{\infty} \mathrm{d}\tau \, e^{i
\, \omega^{\prime} \tau}
\langle B_{j}(\tau) B_{j^{\prime}}(0) \rangle_{\text{th}} \nonumber \\
&= \frac{1}{\hbar^2} \delta_{j j^\prime} \sum_k |g_{j,k}|^2 \Big(
\big(1 + N_{j}(\omega_{j,k})\big) \int_{- \infty}^{\infty}
\mathrm{d}\tau \, e^{i \, (\omega - \omega_{j,k}) \, \tau} +
N_{j}(\omega_{j,k}) \int_{- \infty}^{\infty} \mathrm{d}\tau
\, e^{i \, (\omega + \omega_{j,k}) \, \tau}\Big) \nonumber \\
&= \frac{1}{\hbar^2} \delta_{j j^\prime} \sum_k |g_{j,k}|^2 \Big( 2
\pi \delta(\omega - \omega_{j,k}) \left(1 +
N_{j}(\omega_{j,k})\right) + 2 \pi \delta(\omega + \omega_{j,k})
N_{j}(\omega_{j,k}) \Big)
\nonumber \\
&= \,\, \frac{2}{\hbar} \,\, \delta_{j j^\prime} \int_{0}^{\infty}
\mathrm{d}\omega^{\prime} J_{j}(\omega^{\prime}) \Big( \big(1 +
N_{j}(\omega^{\prime})\big) \delta(\omega - \omega^{\prime}) +
N_{j}(\omega^{\prime}) \delta(\omega + \omega^{\prime})\Big)
\nonumber \\
&=  \,\, \frac{2}{\hbar} \,\, \delta_{j j^\prime}
\begin{cases}
J_{j}(\omega) \big(1 + N_{j}(\omega)\big)
& \mbox{for} \quad \;\;\; 0 < \omega < \infty \\
J_{j}(- \omega) \, N_{j}(- \omega) & \mbox{for} \,\,\,\, - \infty <
\omega < 0
\end{cases} \\
&\equiv \delta_{j j^\prime} \gamma_{j}(\omega) , \nonumber
\end{align}
where the sum over the absolute square of the discrete coupling
constants $g_{j,k}$ is replaced by an integral over a continuous
function $J_j(\omega)$ that is defined as $\pi/\hbar \sum_k
|g_{j,k}|^2 \delta(\omega - \omega_{j,k})$ and called the spectral
density function. This function encapsulates all the effects of the
$j$th bath on the associated pseudo-spin.

Note that $- N_{j}(- \omega)$ equals to $1 + N_{j}(\omega)$. Hence,
if $J_j(\omega)$ is an odd function, $\gamma_{j j}(\omega) \equiv
\gamma_{j}(\omega)$ turns out to be $2/\hbar \, J_j(\omega) \big(1 +
N_{j}(\omega)\big)$ for all values of $\omega$. Also note that
$\gamma_{j j^\prime}(\omega)$ is reduced to $\gamma_{j}(\omega)$
above because each pseudo-spin is associated to an independent
environment. This is expected for the imaginary part of one-sided
Fourier transforms of $\langle B_{j}^{\dagger}(t) B_{j^\prime}(0)
\rangle_{\text{th}}$ as well, i.e., $S_{j j^\prime}(\omega) =
\frac{1}{2 i}\left( \Gamma_{j j^{\prime}}(\omega) -
\Gamma^*_{j^{\prime} \! j}(\omega)\right) = \delta_{j j^\prime} S_{j
j}(\omega) \equiv S_{j}(\omega)$.

\subsubsection{Lamb shift Hamiltonian and dissipator}

To start analyzing the open system dynamics of pseudo-spins,
eigenoperators of the self-Hamiltonian $H_{H\!B}$ should be
calculated using Eq. (\ref{NoiseOp}) with $A_j = \sigma_z^{(j)}$.
Since $H_{H\!B}$ has 4 non-degenerate energy levels, there are
${{4}\choose{2}} = 12$ different transitions in the system. Each
possible nonzero value of Bohr frequency $\omega$ corresponds to one
of these transitions. However, an interaction with the environment
does not need to give rise to a transition always. Hence, to account
for such situations where no transition is enabled, $\omega$ can
take one more value that is equal to zero.

Only $3$ of the $13$ values of $\omega$ correspond to non-zero
eigenoperators, which are
\begin{align} \label{Eq_ComputedNoiseOps}
A_{j}(0) &= - \, | e_3 \rangle\langle e_3 | + | e_4 \rangle\langle
e_4 | , \nonumber \\
A_{j}(\omega_{12}) &= (- 1)^{j} | e_2 \rangle\langle e_1 | ,
\\
A_{j}(\omega_{21}) &= (- 1)^{j} | e_1 \rangle\langle e_2 | .
\nonumber
\end{align}

Then, the Lamb shift Hamiltonian $H_{L\!S}$ becomes $H_{L\!S}^{(1)}
+ H_{L\!S}^{(2)} $ such that
\begin{eqnarray} \begin{aligned} \label{Eq_ComputedHls}
H_{L\!S}^{(j)} = S_j^{\,0} \big( | e_3 \rangle\langle e_3 | + | e_4
\rangle\langle e_4 | \big) &+ S_j^{\,1,2} | e_1 \rangle\langle e_1 |
+ S_j^{\,2,1} | e_2 \rangle\langle e_2 |
\end{aligned}
\end{eqnarray}
where $S_{j}^{\,0} = S_j(0)$ and
$S_{j}^{\,j^\prime\!\!,j^{\prime\prime}} = S_j(\omega_{j^\prime \!
j^{\prime\prime}})$. Similarly, the dissipator $\mathcal{D}(\rho)$
is decomposed into two dissipators each of which takes the following
form
\begin{align} \label{Eq_ComputedDissip}
\mathcal{D}^{(j)}[\rho] = &- 2 \gamma_j^{\,0} \big( \wp_{3,4} | e_3
\rangle\langle e_4 |
+ \wp_{4,3} | e_4 \rangle\langle e_3 | \big) \nonumber \\
&- \frac{1}{2}\big( \gamma_j^{\,0} + \gamma_j^{\,1,2} \big) \big(
\wp_{3,1} | e_3 \rangle\langle e_1 | + \wp_{1,3} | e_1
\rangle\langle e_3 | + \wp_{4,1} | e_4 \rangle\langle e_1 |
+ \wp_{1,4} | e_1 \rangle\langle e_4 | \big) \nonumber  \\
&- \frac{1}{2}\big( \gamma_j^{\,0} + \gamma_j^{\,2,1} \big) \big(
\wp_{3,2} | e_3 \rangle\langle e_2 | + \wp_{2,3} | e_2
\rangle\langle e_3 | + \wp_{4,2} | e_4 \rangle\langle e_2 |
+ \wp_{2,4} | e_2 \rangle\langle e_4 | \big) \nonumber  \\
&- \frac{1}{2}\big( \gamma_j^{\,1,2} + \gamma_j^{\,2,1}\big) \big(
\wp_{1,2} | e_1 \rangle\langle e_2 | + \wp_{2,1} | e_2
\rangle\langle e_1 |\big) - \big( \gamma_j^{\,1,2} \wp_{1,1} -
\gamma_j^{\,2,1} \wp_{2,2} \big) \big( | e_1 \rangle\langle e_1 |  -
| e_2 \rangle\langle e_2 | \big)
\end{align}
with $\gamma_{j}^{\,0} = \gamma_j(0)$,
$\gamma_{j}^{\,j^\prime\!\!,j^{\prime\prime}} =
\gamma_j(\omega_{j^\prime \! j^{\prime\prime}})$, and
$\wp_{j,j^\prime} = \wp_{j,j^\prime}(t) \equiv \langle e_j | \,
\rho(t) | e_{j^\prime} \rangle$ are the elements of the pseudo-spin
density matrix in energy eigenbasis $\{|e_j\rangle\}$.

\subsubsection{Exact solution of the master equation}

When we substitute (\ref{Eq_Ham_HB2}), (\ref{Eq_ComputedHls}) and
(\ref{Eq_ComputedDissip}) into Eq. (\ref{MasterEq}), we end up with
a master equation for the system of our interest. It is
straightforward to solve this master equation analytically. Exact
solution in the energy eigenbasis is given by
\begin{align} \label{Eq_ComputedRho1} \wp_{1,1}(t) &=
\frac{\tilde{\gamma}_{2,1}}{\tilde{\gamma}_{1,2}+\tilde{\gamma}_{2,1}}
\big(\wp_{1,1}(0) + \wp_{2,2}(0)\big) + e^{-
(\tilde{\gamma}_{1,2}+\tilde{\gamma}_{2,1}) t}
\big(\frac{\tilde{\gamma}_{1,2}}{\tilde{\gamma}_{1,2}+\tilde{\gamma}_{2,1}}
\wp_{1,1}(0) -
\frac{\tilde{\gamma}_{2,1}}{\tilde{\gamma}_{1,2}+\tilde{\gamma}_{2,1}}
\wp_{2,2}(0) \big) , \nonumber \\
\wp_{1,2}(t) &= e^{- i \big(\tilde{S}_{1,2} - \tilde{S}_{2,1} +
\omega_{12} \big) t } e^{- \frac{1}{2} \big(\tilde{\gamma}_{1,2}
+ \tilde{\gamma}_{2,1} \big) t} \wp_{1,2}(0) , \nonumber \\
\wp_{1,3}(t) &= e^{+ i \big(\tilde{S}_0 -\tilde{S}_{1,2} +
\omega_{31} \big) t} e^{- \frac{1}{2} \big( \tilde{\gamma}_0  +
\tilde{\gamma}_{1,2} \big) t}
\wp_{1,3}(0) , \nonumber \\
\wp_{1,4}(t) &= e^{+ i \big(\tilde{S}_0 -\tilde{S}_{1,2} +
\omega_{41} \big) t} e^{- \frac{1}{2} \big( \tilde{\gamma}_0  +
\tilde{\gamma}_{1,2} \big) t}
\wp_{1,4}(0) , \nonumber \\
\wp_{2,2}(t) &=
\frac{\tilde{\gamma}_{1,2}}{\tilde{\gamma}_{1,2}+\tilde{\gamma}_{2,1}}
\big(\wp_{1,1}(0) + \wp_{2,2}(0)\big) - e^{-
(\tilde{\gamma}_{1,2}+\tilde{\gamma}_{2,1}) t}
\big(\frac{\tilde{\gamma}_{1,2}}{\tilde{\gamma}_{1,2}+\tilde{\gamma}_{2,1}}
\wp_{1,1}(0) -
\frac{\tilde{\gamma}_{2,1}}{\tilde{\gamma}_{1,2}+\tilde{\gamma}_{2,1}}
\wp_{2,2}(0) \big) , \nonumber \\
\wp_{2,3}(t) &= e^{+ i \big(\tilde{S}_0 -\tilde{S}_{2,1} +
\omega_{32} \big) t} e^{- \frac{1}{2} \big(\tilde{\gamma}_0 +
\tilde{\gamma}_{2,1} \big) t}
\wp_{2,3}(0) , \nonumber \\
\wp_{2,4}(t) &= e^{+ i \big(\tilde{S}_0 -\tilde{S}_{2,1} +
\omega_{42} \big) t} e^{- \frac{1}{2} \big(\tilde{\gamma}_0 +
\tilde{\gamma}_{2,1} \big) t}
\wp_{2,3}(0) , \nonumber \\
\wp_{3,3}(t) &= \wp_{3,3}(0) , \nonumber \\
\wp_{3,4}(t) &= e^{- i \omega_{34} t} e^{- 2 \tilde{\gamma}_0 t}
\wp_{3,4}(0) , \nonumber \\
\wp_{4,4}(t) &= \wp_{4,4}(0) ,
\end{align}
where $\tilde{S}_{0} = S_1^{\,0} + S_2^{\,0}$,
$\tilde{S}_{j^\prime\!\!,j^{\prime\prime}} = S_1^{\,j^\prime \!\!,
j^{\prime\prime}} + S_2^{\,j^\prime \!\!,j^{\prime\prime}}$,
$\tilde{\gamma}_{0} = \gamma_1^{\,0} + \gamma_2^{\,0}$, and
$\tilde{\gamma}_{j^\prime\!\!,j^{\prime\prime}} =
\gamma_1^{\,j^\prime \!\!, j^{\prime\prime}} + \gamma_2^{\,j^\prime
\!\!, j^{\prime\prime}}$.

\subsubsection{Steady state of the master equation}

$\wp_{3,3}$ and $\wp_{4,4}$ are found to be constants of the open
system dynamics in (\ref{Eq_ComputedRho1}). Besides this,
$\wp_{1,1}$ and $\wp_{2,2}$ seem to go to nonzero constant values as
well in the asymptotic limit. On the other hand, all the other
elements of density matrix vanish when $t$ goes to infinity. Let's
show it more clearly by checking the stationary state that is
obtained by taking the left-hand side of master equation given in
(2.7) as zero:
\begin{eqnarray} \label{Eq_ComputedRhoInf0}  \begin{aligned}
\rho^{\infty} &= \wp_{3,3}(0) | e_3 \rangle\langle e_3 | +
\wp_{4,4}(0) | e_4 \rangle\langle e_4 | + \frac{1 - \wp_{3,3}(0) -
\wp_{4,4}(0)}{\tilde{\gamma}_{1,2} + \tilde{\gamma}_{2,1}}
\big(\tilde{\gamma}_{2,1} | e_1 \rangle\langle e_1 | +
\tilde{\gamma}_{1,2} | e_2 \rangle\langle e_2 |\big) .
\end{aligned}
\end{eqnarray}

To elaborate on this calculation, we need to find
$\tilde{\gamma}_{1,2}/(\tilde{\gamma}_{1,2} + \tilde{\gamma}_{2,1})$
and $\tilde{\gamma}_{2,1}/(\tilde{\gamma}_{1,2} +
\tilde{\gamma}_{2,1})$. We can evaluate them for two baths at the
same temperature, e.g, $N_1(\omega) = N_2(\omega) = N(\omega)$ by
using (\ref{Eq_Rates2}) together with the fact that $\omega_{12} = -
\omega_{21} < 0$:
\begin{align}
\frac{\tilde{\gamma}_{1,2}}{\tilde{\gamma}_{1,2}+\tilde{\gamma}_{2,1}}
&= \frac{N(\omega_{21})}{N(\omega_{21}) + \big(1 +
N(\omega_{21})\big)} = \frac{1}{1 + e^{+ \beta (e_2 - e_1)}} =
\frac{e^{- \beta e_2}}{e^{-
\beta e_2} + e^{- \beta e_1}} , \label{Eq_Rates3_1} \\ \nonumber \\
\frac{\tilde{\gamma}_{2,1}}{\tilde{\gamma}_{1,2}+\tilde{\gamma}_{2,1}}
&= \frac{\big(1 + N(\omega_{21})\big)}{N(\omega_{21}) + \big(1 +
N(\omega_{21})\big)} = \frac{e^{+ \beta (e_2 - e_1)}}{1 + e^{+ \beta
(e_2 - e_1)}} = \frac{e^{- \beta \epsilon_1}}{e^{- \beta \epsilon_2}
+ e^{- \beta \epsilon_1}} . \label{Eq_Rates3_2}
\end{align}

Then, the steady state solution given in (\ref{Eq_ComputedRhoInf0})
can be cast into the following simple form:
\begin{eqnarray} \label{Eq_ComputedRhoInf1}  \begin{aligned}
\rho^{\infty} = \wp_{3,3}(0) | e_3 \rangle\langle e_3 | +
\wp_{4,4}(0) | e_4 \rangle\langle e_4 | + \frac{1 - \wp_{3,3}(0) -
\wp_{4,4}(0)}{e^{- \beta e_1} + e^{- \beta e_2}} \big(e^{- \beta
e_1} | e_1 \rangle\langle e_1 | + e^{- \beta e_2} | e_2
\rangle\langle e_2 |\big) .
\end{aligned}
\end{eqnarray}

For the initial states satisfying $\wp_{3,3}(0) =  e^{- \beta e_3} /
\sum_i e^{- \beta e_i}$ and $\wp_{4,4}(0) = e^{- \beta e_4} / \sum_i
e^{- \beta e_i}$, this stationary state turns out to be the thermal
state. However, it doesn't mean that thermalization is the
underlying mechanism for this result. Actually, a partial dephasing
appears to be in charge: environment washes out all the coherence in
the basis of $\{ | e_3 \rangle, | e_4 \rangle \}$, while it imposes
a detailed balance between $| e_1 \rangle$ and $| e_2 \rangle$. As
none of the eigenoperators of $H_{H\!B}$ that corresponds to a
transition from and/or to $e_3$ or $e_4$ survives in
(\ref{Eq_ComputedNoiseOps}), environment can only exchange
information with these two energy levels and this results in a
partial dephasing in the associated energy eigenstates. On the other
hand, the same environment can exchange heat with the remaining
energy levels since there are non-zero eigenoperators for these
transitions and so, it equilibrates energy eigenstates $| e_1
\rangle$ and $| e_2 \rangle$.

In the meantime, note that this two-qubit steady state shares
exactly the same form with the twelve-qubit steady state given in
Eq. (\ref{roInfHex}).

\subsection{Nonlocal proton-phonon coupling} \label{Sec_OpenHBModel_II}

We will extend the open system dynamics to include the oscillations
of O$-$O separation $R_{12}$ in what follows. Assume that
$\hat{u}_j$ is the displacement of the $j$th O atom from its
reference position. Then the deviation of $R_{12}$ from its
equilibrium value $R^{\text{eq}}_{12}$ can be defined as $\Delta
R_{12} = \hat{u}_2 - \hat{u}_1$. By considering this, let's expand
the hopping constant $J_{12}$ about the point $R_{12} =
R^{\text{eq}}_{12}$:
\begin{eqnarray} \begin{aligned} \label{Eq_J_2}
J_{12}(R_{12}) &\approx J_{12}(R^{\text{eq}}_{12}) + \frac{\partial
J_{12}}{\partial R_{12}} \!\!\!\!\!\!\!\underset{R_{12} =
R^{\text{eq}}_{12}}{\Bigg|} \!\!\!\! (\hat{u}_2 - \hat{u}_1) \equiv
J^{(0)}_{12}(R^{\text{eq}}_{12}) + J^{(1)}_{12}(R^{\text{eq}}_{12})
(\hat{u}_2 - \hat{u}_1) .
\end{aligned} \end{eqnarray}

To reduce in complexity and extent, we assume that the first O atom
is stationary, i.e., $\hat{u}_1 = 0$. Then, we switch into the
second-quantization representation of $\hat{u}_2$ replacing it with
$\sum_k \sqrt{\hbar / (2 \mu \Omega_{k})}(d^{\dagger}_{2,k} +
d_{2,k})$ where $\Omega_k$ are the frequencies of the oscillation of
$R_{12}$, and $d^{\dagger}_{2,k}$ and $d_{2,k}$ are respectively the
phonon creation and annihilation operators associated with the
vibration of the second O atom. After this replacement, substitution
of (\ref{Eq_J_2}) into (\ref{Eq_Ham_HB2}) causes the transformation
$H_{H\!B} \rightarrow H_{H\!B} + H^{nl}_I$ where the value of
parameter $J_x$ in $H_{H\!B}$ turns out to be $-
J^{(0)}_{12}(R^{\text{eq}}_{12})/2$ and the $H^{nl}_I$ is a nonlocal
proton-phonon interaction described by
\begin{eqnarray} \begin{aligned} \label{Eq_HamI2}
H^{nl}_I &= \left(\sigma^{(1)}_{x} \otimes \sigma^{(2)}_{x} +
\sigma^{(1)}_{y} \otimes \sigma^{(2)}_{y} \right) \sum_k  \, h_k
\left(d_{2,
k}^{\dagger} + d_{2, k}\right) \\
&\equiv A_3 \otimes B_3 ,
\end{aligned} \end{eqnarray}
with $h_k$ equals to $- J^{(1)}_{12}(R^{\text{eq}}_{12})
\sqrt{\hbar/(8 \mu \Omega_{k})}$. This new proton-phonon interaction
requires to entail the calculation of one more non-zero
eigenoperators of $H_{H\!B}$:
\begin{align} \label{Eq_ComputedNoiseOps_Add}
A_{3}(0) &= 2 (| e_1 \rangle\langle e_1 | - | e_2 \rangle\langle e_2
|) ,
\end{align}
that give rise to the emergence of the following Lamb-shift
Hamiltonian in addition to the ones given in (\ref{Eq_ComputedHls}):
\begin{eqnarray} \begin{aligned} \label{Eq_ComputedHls_Add}
H_{L\!S}^{(3)} = 4 S_3^{\,0} \left( | e_1 \rangle\langle e_1 | + |
e_2 \rangle\langle e_2 | \right) ,
\end{aligned}
\end{eqnarray}
and the following dissipator in addition to the ones given in
(\ref{Eq_ComputedDissip}):
\begin{align} \label{Eq_ComputedDissip_Add}
\mathcal{D}^{(3)}[\rho] = &- 2 \gamma_3^{\,0} \big( \wp_{3,1} | e_3
\rangle\langle e_1 | + \wp_{1,3} | e_1 \rangle\langle e_3 | +
\wp_{4,1} | e_4 \rangle\langle e_1 |
+ \wp_{1,4} | e_1 \rangle\langle e_4 | \big) \nonumber  \\
&- 2 \gamma_3^{\,0} \big( \wp_{3,2} | e_3 \rangle\langle e_2 | +
\wp_{2,3} | e_2 \rangle\langle e_3 | + \wp_{4,2} | e_4
\rangle\langle e_2 |
+ \wp_{2,4} | e_2 \rangle\langle e_4 | \big) \nonumber  \\
&- 8 \gamma_3^{\,0} \big( \wp_{1,2} | e_1 \rangle\langle e_2 | +
\wp_{2,1} | e_2 \rangle\langle e_1 |\big) .
\end{align}

Inclusion of these additional terms into the master equation changes
the exact solution from (\ref{Eq_ComputedRho1}) to:
\begin{align} \label{Eq_ComputedRho2} \wp_{1,1}(t) &=
\frac{\tilde{\gamma}_{2,1}}{\tilde{\gamma}_{1,2}+\tilde{\gamma}_{2,1}}
\big(\wp_{1,1}(0) + \wp_{2,2}(0)\big) + e^{-
(\tilde{\gamma}_{1,2}+\tilde{\gamma}_{2,1}) t}
\big(\frac{\tilde{\gamma}_{1,2}}{\tilde{\gamma}_{1,2}+\tilde{\gamma}_{2,1}}
\wp_{1,1}(0) -
\frac{\tilde{\gamma}_{2,1}}{\tilde{\gamma}_{1,2}+\tilde{\gamma}_{2,1}}
\wp_{2,2}(0) \big) , \nonumber \\
\wp_{1,2}(t) &= e^{- i \big(\tilde{S}_{1,2} - \tilde{S}_{2,1} +
\omega_{12} \big) t } e^{- \frac{1}{2} \big(\tilde{\gamma}_{1,2}
+ \tilde{\gamma}_{2,1} + 16 \gamma_3^{\,0} \big) t} \wp_{1,2}(0) , \nonumber \\
\wp_{1,3}(t) &= e^{+ i \big(\tilde{S}_0 -\tilde{S}_{1,2} - 4
S_3^{\,0} + \omega_{31} \big) t} e^{- \frac{1}{2}
\big(\tilde{\gamma}_0 + \tilde{\gamma}_{1,2} + 4 \gamma_3^{\,0}
\big) t}
\wp_{1,3}(0) , \nonumber \\
\wp_{1,4}(t) &= e^{+ i \big(\tilde{S}_0 -\tilde{S}_{1,2} - 4
S_3^{\,0} + \omega_{41} \big) t} e^{- \frac{1}{2}
\big(\tilde{\gamma}_0 + \tilde{\gamma}_{1,2} + 4 \gamma_3^{\,0}
\big) t}
\wp_{1,4}(0) , \nonumber \\
\wp_{2,2}(t) &=
\frac{\tilde{\gamma}_{1,2}}{\tilde{\gamma}_{1,2}+\tilde{\gamma}_{2,1}}
\big(\wp_{1,1}(0) + \wp_{2,2}(0)\big) - e^{-
(\tilde{\gamma}_{1,2}+\tilde{\gamma}_{2,1}) t}
\big(\frac{\tilde{\gamma}_{1,2}}{\tilde{\gamma}_{1,2}+\tilde{\gamma}_{2,1}}
\wp_{1,1}(0) -
\frac{\tilde{\gamma}_{2,1}}{\tilde{\gamma}_{1,2}+\tilde{\gamma}_{2,1}}
\wp_{2,2}(0) \big) , \nonumber \\
\wp_{2,3}(t) &= e^{+ i \big(\tilde{S}_0 -\tilde{S}_{2,1} - 4
S_3^{\,0} + \omega_{32} \big) t} e^{- \frac{1}{2}
\big(\tilde{\gamma}_0 + \tilde{\gamma}_{2,1} + 4 \gamma_3^{\,0}
\big) t}
\wp_{2,3}(0) , \nonumber \\
\wp_{2,4}(t) &= e^{+ i \big(\tilde{S}_0 -\tilde{S}_{2,1} - 4
S_3^{\,0} + \omega_{42} \big) t} e^{- \frac{1}{2}
\big(\tilde{\gamma}_0 + \tilde{\gamma}_{2,1} + 4 \gamma_3^{\,0}
\big) t}
\wp_{2,3}(0) , \nonumber \\
\wp_{3,3}(t) &= \wp_{3,3}(0) , \nonumber \\
\wp_{3,4}(t) &= e^{- i \omega_{34} t} e^{- 2 \tilde{\gamma}_0 t}
\wp_{3,4}(0) , \nonumber \\
\wp_{4,4}(t) &= \wp_{4,4}(0) .
\end{align}

On the other hand, since the diagonal elements have no dependence on
either $\gamma_3^{\,0}$ or $S_3^{\,0}$, the stationary state remains
the same as
\begin{eqnarray} \label{Eq_ComputedRhoInf2}  \begin{aligned}
\rho^{\infty} &= \wp_{3,3}(0) | e_3 \rangle\langle e_3 | +
\wp_{4,4}(0) | e_4 \rangle\langle e_4 | + \frac{1 - \wp_{3,3}(0) -
\wp_{4,4}(0)}{\tilde{\gamma}_{1,2} + \tilde{\gamma}_{2,1}}
\big(\tilde{\gamma}_{2,1} | e_1 \rangle\langle
e_1 | + \tilde{\gamma}_{1,2} | e_2 \rangle\langle e_2 |\big) \\
&= \wp_{3,3}(0) | e_3 \rangle\langle e_3 | + \wp_{4,4}(0) | e_4
\rangle\langle e_4 | + \frac{1 - \wp_{3,3}(0) - \wp_{4,4}(0)}{e^{-
\beta e_1} + e^{- \beta e_2}} \big(e^{- \beta e_1} | e_1
\rangle\langle e_1 | + e^{- \beta e_2} | e_2 \rangle\langle e_2
|\big) .
\end{aligned}
\end{eqnarray}

In this respect, O$-$O vibrations change the dynamics of the system,
but do not affect its steady state.

\newpage

\section{Model Parameters}

\subsection{Estimation of the parameters} \label{II-A}

Note that if the twelve-qubit initial state $\rho(t=0)$ lives only
in the $64$-dimensional subspace $\mathcal{H}_{ice}$, the steady
state of the chosen master equation depends on two free parameters,
$J_x$ and $J_z^{(\text{intra})}$. Here, $J_x$ equals to half of the
orbital interaction energy $J$ which is responsible for the
tunneling of the protons between O atoms, while
$J_z^{(\text{intra})}$ is a quarter of the inter-proton interaction
energy $V_{intra}$ which is responsible for the ionic defect
penalty.

The values of these free parameters are extracted comparing the
temperature dependent behaviour of probability $P_{BF}$ with the
phase transition temperatures predicted by recent dielectric
constant measurements \cite{2015_HexPhaseTransition,
2015_CorrPTInIceXI} as follows.

\begin{figure}[h] \centering
        \includegraphics[width=0.57\textwidth]{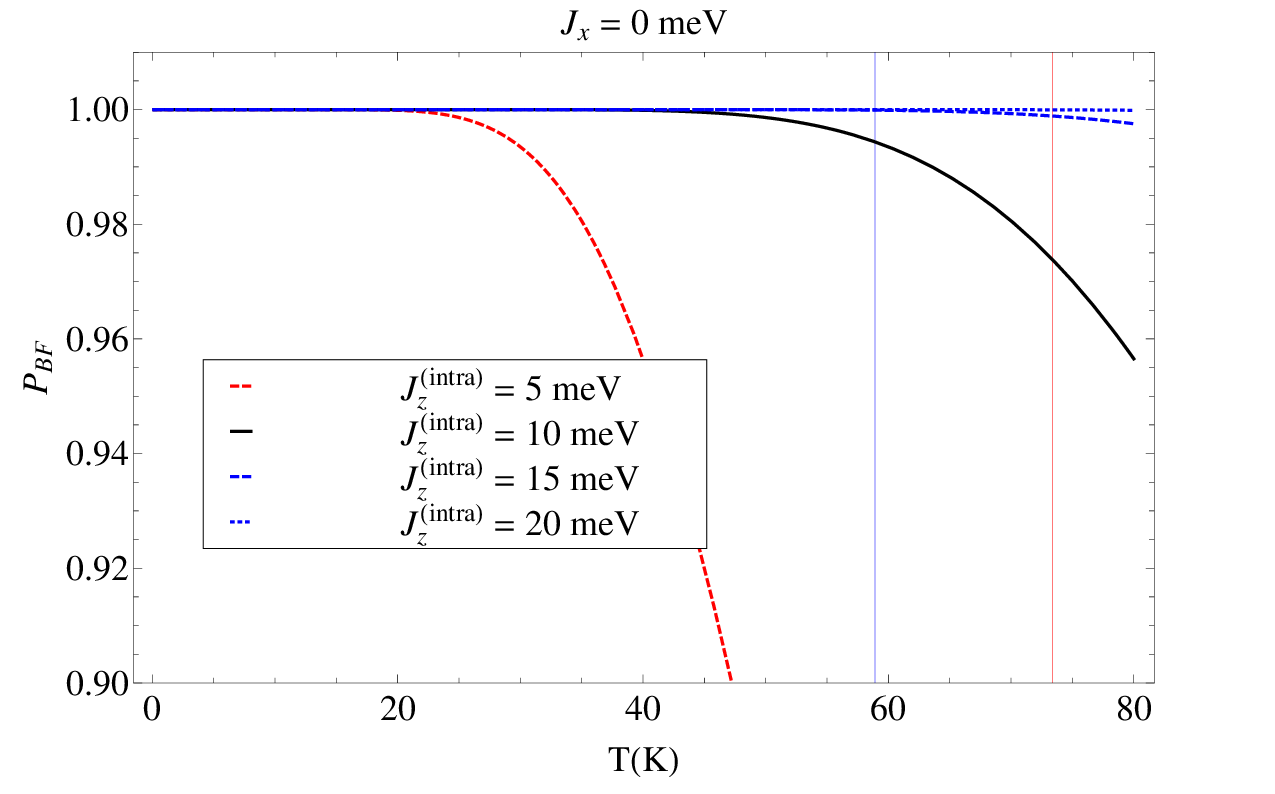}
        \caption{The behaviour of probability $P_{BF}(T)$ with respect to
        changes in the energy of ionic defect penalty
        when the orbital interaction energy vanishes.
        Vertical solid lines coloured blue and red respectively pinpoint the experimentally determined
        phase transition temperatures $58.9$ K and $73.4$ K \cite{2015_HexPhaseTransition}.}
        \label{Fig_PWithJxEq0}
\end{figure}

First, we set $J_x$ to zero and search for the appropriate
$J_z^{(\text{intra})}$ values that give the expected temperature
dependence of $P_{BF}$, i.e., $P_{BF}(T)$ should be sufficiently
close to unity at temperatures lower than the experimentally
determined phase transition temperatures and show a decrease during
the phase transition. In this way, we try to reproduce the
experimental data without any need to assume that proton tunneling
takes place during the phase transition. As shown in Table
\ref{Table_JxEq0} and Fig. \ref{Fig_PWithJxEq0}, $10$ meV is the
maximum value consistent with the experimental data when compared to
its close neighborhood. We set $J_z^{(\text{intra})}$ to $10$ meV in
this respect.

Secondly, we gradually decrease $J_x$ and search for its minimum
value that preserves the consistency with the experimental data. The
value of $J_x$ found in this way is $- 0.5$ meV as shown in Table
\ref{Table_Jx}.

Note that this two-step procedure does not exclude the likelihood of
the presence of any other $(J_z^{(\text{intra})},J_x)$ pair in the
phase space that might lead to exactly the same $P_{BF}(T)$ as shown
in Fig. \ref{SandP} in the manuscript. However, it offers a
physically motivated $(J_z^{(\text{intra})},J_x)$ pair as described
above.

\subsection{Sensitivity of the parameters} \label{II-B}

The sensitivity of $P_{BF}(T)$ to the changes in the free parameters
$J_x$ and $J_z^{(\text{intra})}$ will be investigated in what
follows.

\begin{figure}[h] \centering
        \includegraphics[width=0.57\textwidth]{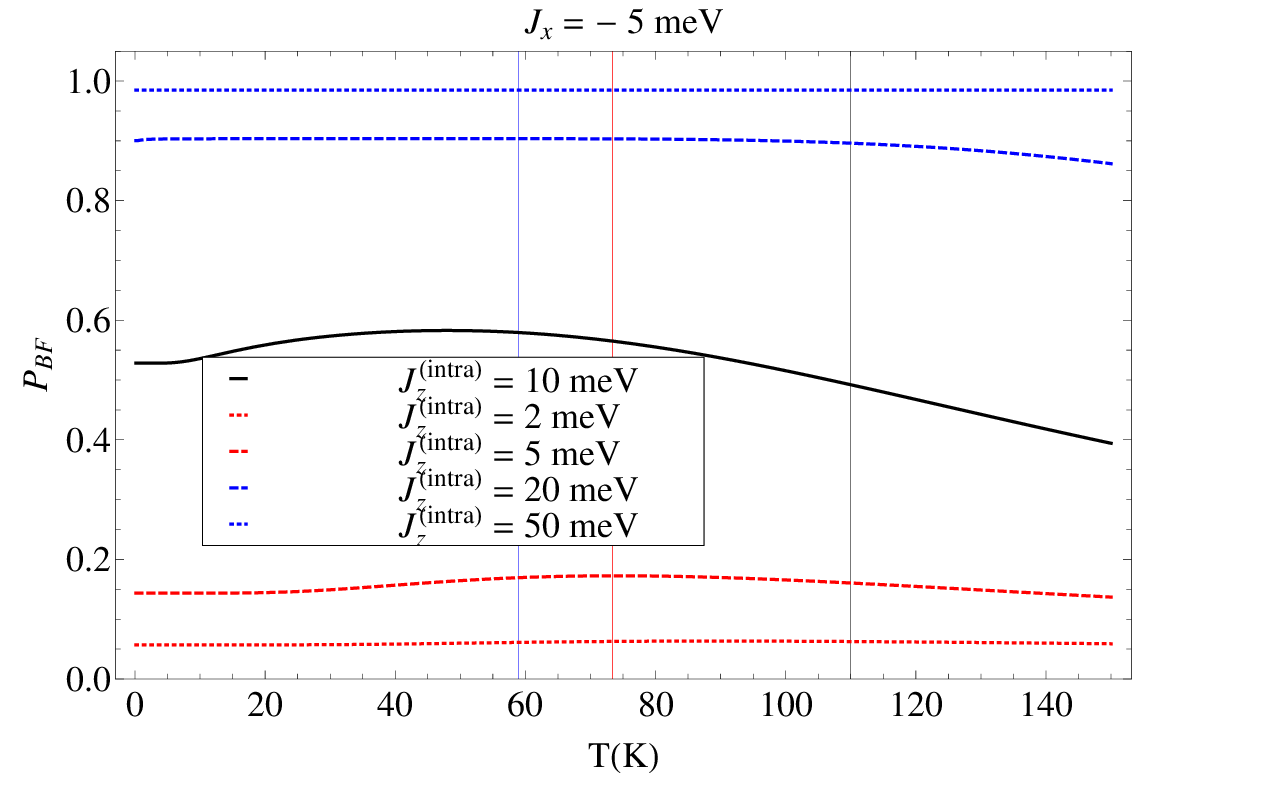}
        \caption{The behaviour of probability $P_{BF}(T)$ with respect to
        further changes in the energy of ionic defect penalty
        after the orbital interaction energy is reduced to a value lower than
        the fixed value used in the manuscript.}
        \label{Fig_PWithJxEq5}
\end{figure}

Although $P_{BF}(T)$ should be sufficiently close to unity at low
temperatures, Fig. \ref{{Pv2}}-a given in the manuscript shows that
it cannot reach to this limit at any temperature when
$J_z^{(\text{intra})}$ is kept constant at $+ 10$ meV but $J_x$ is
set to a value less than $- 0.5$ meV, e.g. to $-5$ meV. Furthermore,
Fig. \ref{Fig_PWithJxEq5} given below displays that it is impossible
to readjust the value of $J_z^{(\text{intra})}$ to bring the
temperature dependence of $P_{BF}$ back to the expected behaviour
after decreasing $J_x$ down to $-5$ meV. In fact, some values of
$J_z^{(\text{intra})}$ are found to raise $P_{BF}$ up to unity at
low temperatures, but $P_{BF}(T)$ never decreases for these
particular $J_z^{(\text{intra})}$ values, even at temperatures quite
higher than the experimentally determined phase transition
temperatures. Note that $P_{BF}(T)$ should show a decrease during
the phase transition. Hence, an increase in the proton tunneling
rate, up to a value ten times higher than the fixed value used in
the manuscript, cannot be compensated by a further change in the
energy of ionic defect penalty.

\begin{figure}[h] \centering
        \includegraphics[width=0.57\textwidth]{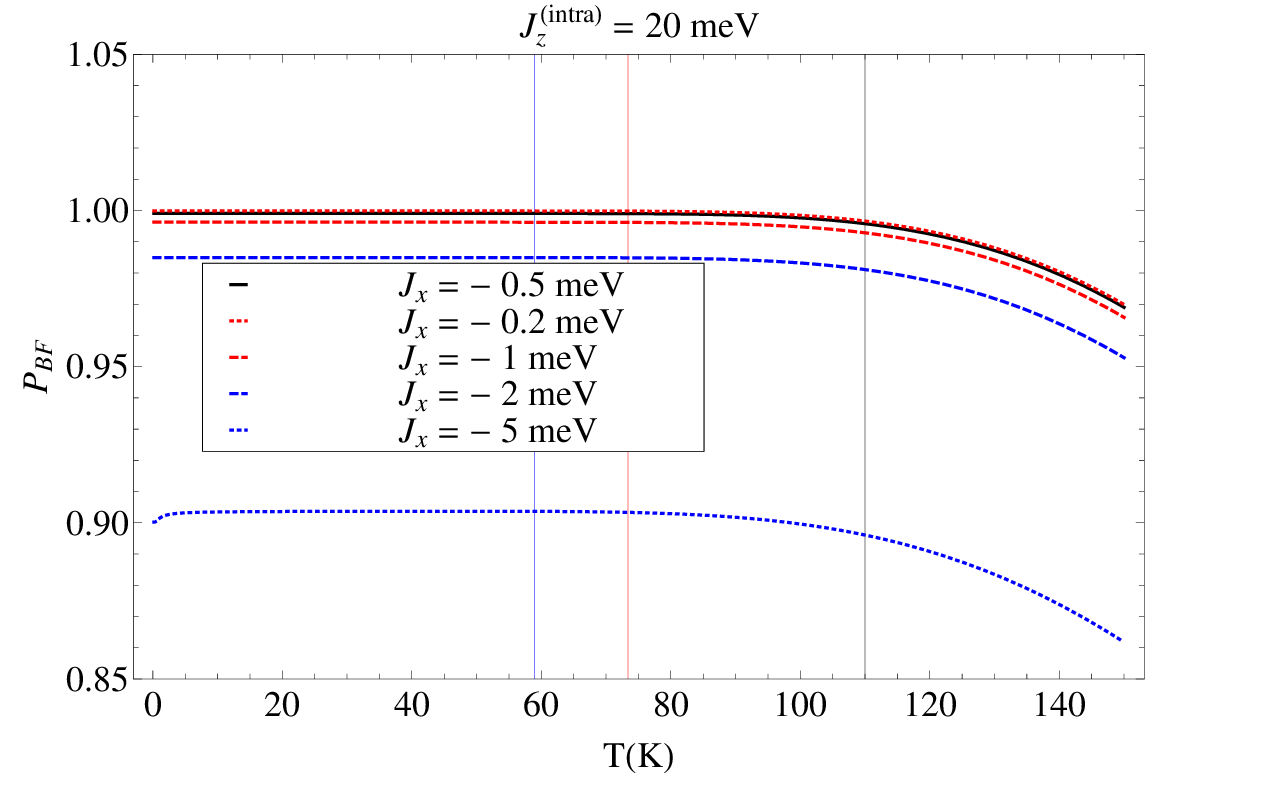}
        \caption{The behaviour of probability $P_{BF}(T)$ with respect to
        further changes in the orbital interaction energy
        after the energy of ionic defect penalty is raised to a value
        higher than the fixed value used in the manuscript.}
        \label{Fig_PWithJzEq20}
\end{figure}

\begin{figure}[h] \centering
        \includegraphics[width=0.57\textwidth]{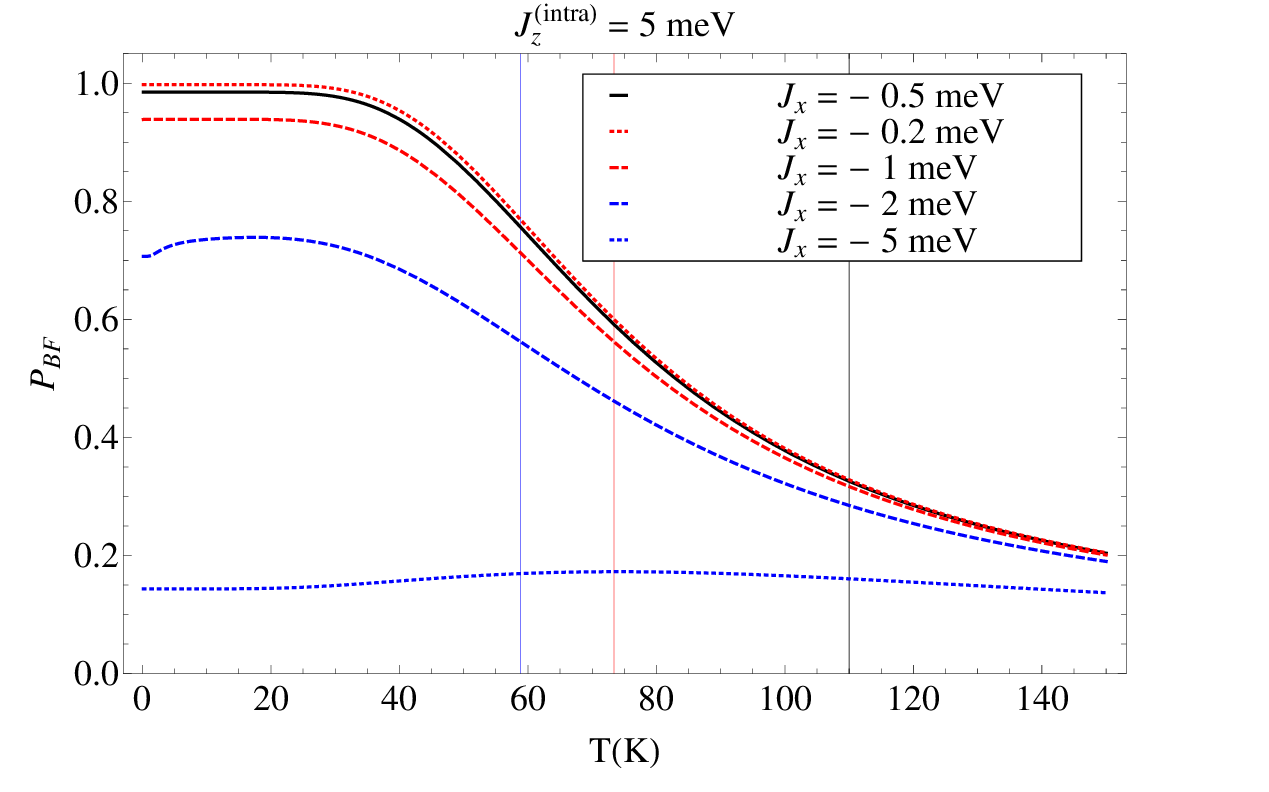}
        \caption{The behaviour of probability $P_{BF}(T)$ with respect to
        further changes in the orbital interaction energy
        after the energy of ionic defect penalty is reduced to
        a value lower than the fixed value used in the manuscript.}
        \label{Fig_PWithJzEq5}
\end{figure}

On the other hand, according to Fig. \ref{Pv2}-b given in the
manuscript, a change in the value of $J_z^{(\text{intra})}$ from
$10$ meV to $20$ meV ($5$ meV) sets the temperature at which
$P_{BF}(T)$ deviates from unity to a value higher (lower) than the
experimentally determined phase transition temperatures. Also,
$P_{BF}(T)$ fails to exhibit its proper behaviour after (before)
this turning point when compared to Fig. \ref{SandP} given in the
manuscript. Here, Fig. \ref{Fig_PWithJzEq20} (Fig.
\ref{Fig_PWithJzEq5}) demonstrates that no further adjustment in
$J_x$ can regenerate the expected temperature dependence of $P_{BF}$
after varying $J_z^{(\text{intra})}$ to $20$ meV ($5$ meV). Hence, a
change in the energy of ionic defect penalty, up to a value twice as
high (low) as the fixed value used in the manuscript, cannot be
neutralized by readjusting the proton tunneling rate.

In this respect, the expected temperature dependence of $P_{BF}$
exhibits a sensitivity to our free parameters, i.e., deviations from
the fixed value of one parameter prevent the appearance of a slow
decline in $P_{BF}(T)$ from unity around $58.9 - 73.4$ K, and this
behaviour cannot reappear when the second parameter is also allowed
to deviate from its fixed value at the same time.
\begin{table}[h!]
\centering \caption{Dependence of $P_{BF}(T)$ and $S(T)$ to
$J_z^{(\text{intra})}$ when $J_x = 0$.}
\begin{tabular}[c]{|c|c|c|c|c|}
\hline & & & &
\\$J_z^{(\text{intra})}$ & $S(20\,\text{K})$ & $P_{BF}(20\,\text{K})$
& $P_{BF}(58.9\,\text{K})$ & $P_{BF}(73.4\,\text{K})$ \\
& & & & \\ \hline & & & & \\
$0.10\,\text{meV}$ & $5.99$ & $0.04$ & $0.04$ & $0.03$ \\ & & & & \\
\hline
& & & & \\
$1.00\,\text{meV}$ & $4.65$ & $0.38$ & $0.09$ & $0.07$ \\ & & & & \\
\hline
& & & & \\
$8.00\,\text{meV}$ & $1.00$ & $1.00$ & $0.97$ & $0.91$ \\ & & & & \\
\hline
& & & & \\
$9.00\,\text{meV}$ & $1.00$ & $1.00$ & $0.99$ & $0.95$ \\ & & & & \\
\hline
& & & & \\
$10.00\,\text{meV}$ & $1.00$ & $1.00$ & $0.99$ & $0.97$ \\ & & & & \\
\hline
& & & & \\
$11.00\,\text{meV}$ & $1.00$ & $1.00$ & $1.00$ & $0.99$ \\ & & & & \\
\hline
& & & & \\
$12.00\,\text{meV}$ & $1.00$ & $1.00$ & $1.00$ & $0.99$ \\ & & & & \\
\hline
& & & & \\
$0.10\,\text{eV}$ & $1.00$ & $1.00$ & $1.00$ & $1.00$ \\ & & & & \\
\hline
& & & & \\
$1.00\,\text{eV}$ & $1.00$ & $1.00$ & $1.00$ & $1.00$ \\ & & & & \\
\hline
\end{tabular}
\label{Table_JxEq0}
\end{table}
\begin{table}[h]
\centering \caption{Dependence of $P_{BF}(T)$ and $S(T)$ to $J_x$
when $J_z^{(\text{intra})} = + 10$ meV.}
\begin{tabular}[c]{|c|c|c|c|c|}
\hline & & & &
\\$J_x(\text{meV})$ & $S(20\,\text{K})$ & $P_{BF}(20\,\text{K})$
& $P_{BF}(58.9\,\text{K})$ & $P_{BF}(73.4\,\text{K})$ \\
& & & & \\ \hline & & & & \\
$- 1.00$ & $1.00$ & $0.98$ & $0.98$ & $0.96$ \\ & & & & \\
\hline
& & & & \\
$- 0.90$ & $1.00$ & $0.99$ & $0.98$ & $0.96$ \\ & & & & \\
\hline
& & & & \\
$- 0.80$ & $1.00$ & $0.99$ & $0.98$ & $0.96$ \\ & & & & \\
\hline
& & & & \\
$-0.70$ & $1.00$ & $0.99$ & $0.99$ & $0.97$ \\ & & & & \\
\hline
& & & & \\
$-0.60$ & $1.00$ & $0.99$ & $0.99$ & $0.97$ \\ & & & & \\
\hline
& & & & \\
$-0.50$ & $1.00$ & $1.00$ & $0.99$ & $0.97$ \\ & & & & \\
\hline
& & & & \\
$-0.40$ & $1.00$ & $1.00$ & $0.99$ & $0.97$ \\ & & & & \\
\hline
& & & & \\
$-0.30$ & $1.00$ & $1.00$ & $0.99$ & $0.97$ \\ & & & & \\
\hline
& & & & \\
$-0.20$ & $1.00$ & $1.00$ & $0.99$ & $0.97$ \\ & & & & \\
\hline
& & & & \\
$-0.10$ & $1.00$ & $1.00$ & $0.99$ & $0.97$ \\ & & & & \\
\hline
& & & & \\
$-0.00$ & $1.00$ & $1.00$ & $0.99$ & $0.97$ \\ & & & & \\
\hline
\end{tabular}
\label{Table_Jx}
\end{table}


\end{document}